\pdfoutput=1
\documentclass[fleqn,usenatbib]{mnras}

\usepackage{newtxtext,newtxmath}
\usepackage{amsmath}

\usepackage[T1]{fontenc}
\usepackage{subcaption}
\usepackage{comment}
\usepackage{makecell}
\usepackage{float}
\usepackage{rotating}
\usepackage{graphicx}	
\usepackage{amsmath}	
\DeclareRobustCommand{\VAN}[3]{#2}
\let\VANthebibliography\thebibliography
\def\thebibliography{\DeclareRobustCommand{\VAN}[3]{##3}\VANthebibliography}

\title[Tailoring intrinsic-alignment priors to galaxy demographics]{Tailoring intrinsic-alignment priors to galaxy demographics for next-generation weak lensing} 

\author[McCullough et al.]{
J. McCullough,$^{1}$\thanks{E-mail: jmccullough@princeton.edu}
J. Siegel,$^{1}$
A. Amon,$^{1}$
B. Joachimi,$^{2}$
D. Gruen,$^{3,4}$
and D. Navarro-Gironés$^5$
\\
$^{1}$Department of Astrophysical Sciences, Princeton University, Peyton Hall, Princeton, NJ 08544, USA\\
$^{2}$Department of Physics and Astronomy, University College London, Gower Street, London WC1E 6BT, UK\\
$^{3}$Universit\"ats-Sternwarte, Fukult\"at f\"ur Physik, Ludwig-Maximilians Universit\"at M\"unchen, Scheinerstr. 1, 81679 M\"unchen, Germany\\
$^{4}$Excellence Cluster ORIGINS, Boltzmannstr. 2, 85748 Garching, Germany \\
$^{5}$Leiden Observatory, Leiden University, Einsteinweg 55, 2333 CC Leiden, The Netherlands
}

\date{Accepted XXX. Received YYY; in original form ZZZ}

\pubyear{\the\year{}}

\begin{document}
\label{firstpage}
\pagerange{\pageref{firstpage}--\pageref{lastpage}}
\maketitle

\begin{abstract}
The intrinsic alignment (IA) of galaxies is a major systematic limiting the precision of weak gravitational lensing cosmology.  
Although direct IA measurements now span increasingly diverse galaxy populations and reveal strong dependence on physical properties, cosmological analyses have typically adopted broad, survey-wide priors that do not reflect the demographics of the galaxies being analyzed. We propose an empirical approach to tailor the IA prior to the lensing galaxy sample according to their overlap with a library of existing direct measurements across a flexible number of astrophysical properties (in this work, stellar mass and specific star-formation rate). 
The method explicitly accounts for incomplete observational coverage, retaining broad priors for populations that lack direct calibration, and naturally extends to future measurements and different IA models. 
We demonstrate the framework using existing direct IA measurements interpreted with the nonlinear alignment (NLA) model to derive priors for selections of galaxies in the Dark Energy Survey COSMOS deep field. We incorporate the method into infrastructure used for photometric redshift calibration (SOMPZ), providing a pathway toward jointly informed redshift and IA calibration in future surveys. 
We find that approximately one-third of the full and blue source samples, but more than half of the red sample, falls outside the effective coverage of existing direct IA measurements --  owing to sparse measurements at low stellar mass. Our approach enables empirical IA priors tailored to individual source samples and offers a practical tool for guiding future direct measurements and optimizing weak-lensing galaxy selection.

\end{abstract}




\begin{figure*}
    \centering 
    \includegraphics[width=0.99\linewidth]{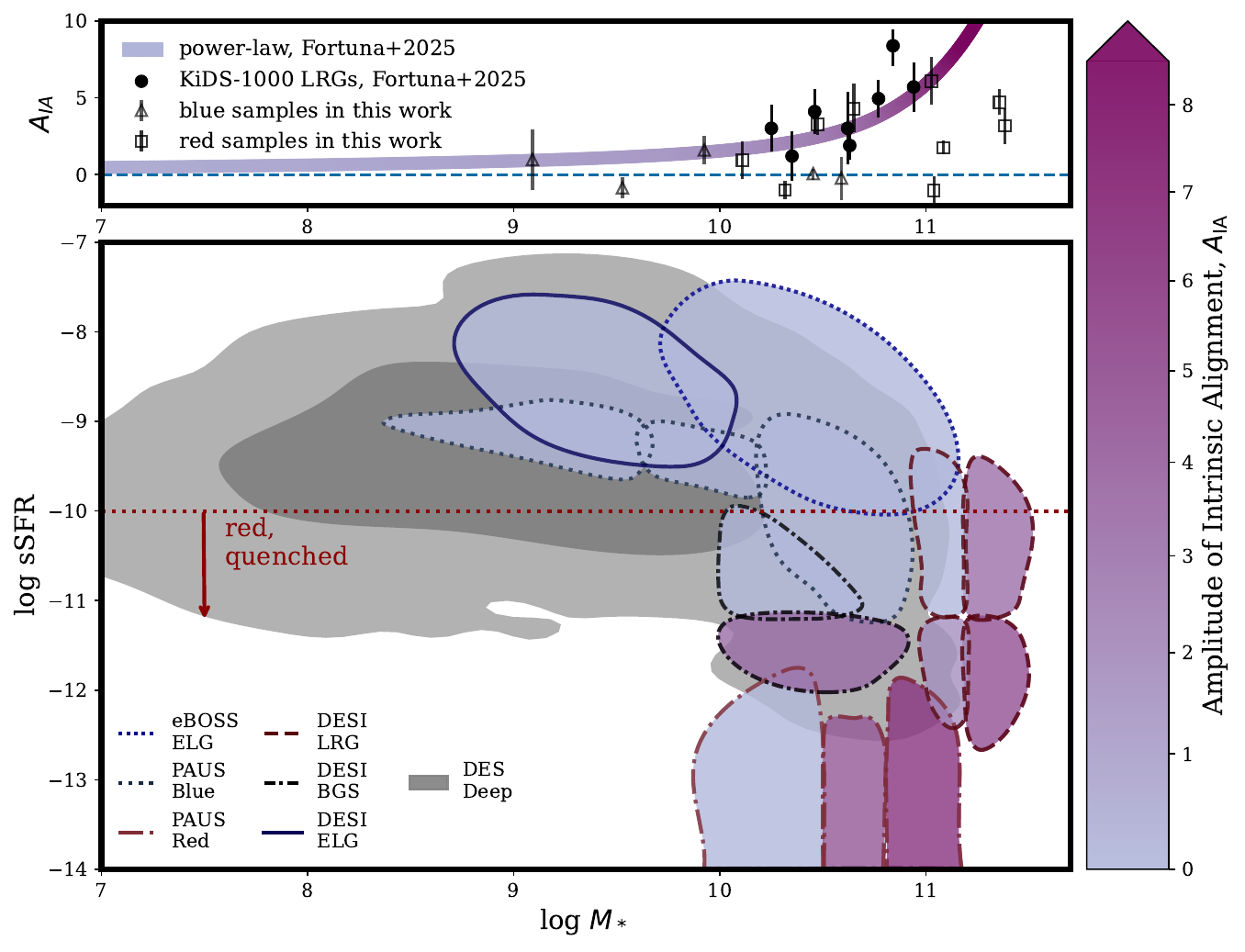}
    \caption{\textbf{(Top)} The single-power-law relation fit in \protect\cite{fortuna_2025} relating galaxy halo-mass to intrinsic alignment, converted to stellar mass for ease of comparison. The data informing this fit (black) are plotted alongside the observations used for this work (unfilled), split between intrinsically blue populations (triangle, e.g., ELGs) and those that contain redder samples (square, e.g., LRGs).\\ 
    \textbf{(Bottom)} The plane of galaxy populations used for IA direct measurements (color, shaded) in specific star formation rate and stellar mass, colored by the amplitude of alignment ($A_{\rm{IA}}$) fit to their respective $w_{g+}$ measurements. Contours are drawn for 68\%-regions of the populations used for inference in this work, which are limited to those with readily available astrophysical properties. Any measurement consistent with zero within 2$\sigma$ is marked as null. Overplotted (dashed, red) is a commonly chosen boundary between passive and star-forming galaxies. The photometric deep COSMOS field population for the Dark Energy Survey \protect\citep{des_deep_field_y3},
    matched to COSMOS 2025 SED fits \protect\citep{cosmos_2025} with estimates on star-formation and mass, is depicted in shaded gray for the 68\%-region (dark gray) and the 95\%-region (light gray). Alignment has been measured to increase as galaxies get redder and more massive, as seen by the amplitudes increasing towards the bottom right corner.}
    \label{fig:diagnost}
\end{figure*}


\section{Introduction}
Weak gravitational lensing has become one of the premier probes of cosmology \citep[for review see e.g., ][]{prat_review, kilbinger_2015, hoekstra_jain_2008, bartelmannschneider_2001}, delivering percent-level constraints on the growth of structure \citep{DES,kids_survey, hsc_survey} and dark energy \citep[e.g.,][]{DESwowa}. The next generation of surveys, including Rubin \citep{lsst_survey,rubin_early_science}, Euclid \citep{euclid_def,euclid_mission} and Roman \citep{roman_overview,roman_survey_science}, will increase the statistical power of existing galaxy shape measurements by an order of magnitude, placing correspondingly tighter requirements on the control of astrophysical systematics. Among these, intrinsic alignments (IA), the coherent alignment of galaxy shapes with the surrounding large-scale structure, remain one of the dominant sources of uncertainty, potentially biasing cosmological inference and limiting the precision achievable with future surveys. \citep[for review see e.g.,][]{chisari_review,lamman_review,troxel_review,joachimi_review}.

Advances in large spectroscopic surveys and multi-band photometric modeling have transformed the observational landscape of direct measurements of IA. With the availability of millions of galaxy spectra from surveys like DESI \citep{desi_y1}, the resulting direct measurements from spectroscopic surveys \citep[e.g.,][]{siegel_2025,hervas_peters_2024_unions,samuroff_2023, johnston_2019,cmass_ia_2013,joachimi_megaz_2011}, many-band photometric redshifts \citep[e.g. PAUS:][]{paus_direct,johnston_2021}, and reliable photometric redshifts from passive, quenched galaxies \citep[e.g.,][]{samuroff_2023,Fortuna_2021} together now supply significant coverage across the astrophysical properties of modern weak-lensing galaxies. This growing library of direct measurements has revealed that IA depends strongly on intrinsic galaxy properties, including if it is star-forming or quiescent and its stellar mass (see Fig.~\ref{fig:diagnost}, Sec.~\ref{sec:library}). Yet, despite previous efforts to inform IA priors using direct measurements, \citep[e.g.,][]{johnston_2019}, this growing empirical knowledge for such diverse galaxy populations has not been systemically translated without extrapolation into IA priors for the specific galaxy populations used in weak-lensing cosmology. In this work, we develop a framework to bridge this gap, using the overlap in astrophysical properties between direct IA measurement samples and weak-lensing source galaxies to construct empirically informed, population-dependent priors.

On the theoretical side, increasingly sophisticated physical models have been developed, extending beyond the nonlinear alignment model \citep{bridle_king_2007} and other tidal alignment approaches \citep[e.g.,][]{blazek_2015} to tidal alignment and tidal torque models \citep[][]{blazek_2019, hirata_seljak_2004}, halo-based descriptions \citep[e.g.,][]{halomodel,fortuna_2021_halomodel, fortuna_2025}, Lagrangian descriptions \citep[e.g.,][]{hymalaia, chen_2024_lagrangian}, and effective field theory approaches \citep{eft,eft_ii,eft_sims,chen_eft}. 
Hydrodynamical simulations have likewise advanced our understanding of the physical origins of alignments and their dependence on galaxy properties \citep[e.g.,][]{horizon_2025, sim_heukelum_cross_sim_disk_ellipticals,euclid_flagship_sim_nla_tatt,euclid_flagship_sim_calib,sim_flamingo_ia,milleniumtng_ia, illustris_tng_ia,samuroff_2021,kokron2026}.
While increasingly flexible models have improved agreement with observations, particularly on small scales ($< 6$ Mpc h$^{-1}$), they introduce additional degrees of freedom that can substantially degrade cosmological precision \citep{chenderose2024}. Conversely, insufficient model flexibility can bias cosmological inference \citep[e.g.,][]{ia_model_specification_bias}. These developments underscore the importance of empirical constraints that can inform the parameters of IA models without relying solely on theoretical assumptions.

Despite these advances in observations, theory, and simulations, cosmological analyses have historically assigned broad IA priors to the full source samples, largely  agnostic to the galaxy population demographics (e.g., \citealt{desy6_cosmic_shear,amon_2021,secco_2021}, \citealt{hsc_real,hsc_fourier}, \citealt{kids1000_cosmic_shear}). Intending to account for variation in aligning populations between tomographic bins, these parameterizations assume strict forms for the redshift evolution of IA; when these assumptions are relaxed, cosmological precision is substantially degraded \citep{eft_desi_des_redshift_evol}. Such flexibility can also allow for interplay with other systematics, particularly photometric redshift uncertainties, potentially biasing results \citep[e.g.,][]{fischbacher_redshift_ia_interplay,ia_photz_degen}. 
Recent analyses have begun incorporating astrophysical information with galaxy property-dependent IA parameterizations (e.g., KiDS-Legacy with NLA-M: \citealt{Wright_2025,kids_legacy_consistency}; UNIONS: \citealt{unions}; TATT-M: \citealt {sim_flamingo_ia}). However, these early efforts have often required extrapolation to substantially lower-mass galaxy populations than those directly measured, relying primarily on alignment measurements of passive galaxies. The growing library of direct IA measurements, together with increasingly detailed characterization of weak-lensing source populations \citep[e.g., ][]{galsbi-sps,galsbi,pop-cosmos,popsed}, now offers an opportunity to construct informative priors that reflect the demographics of the galaxies actually entering cosmological analyses.


A natural framework for doing so already exists in photometric-redshift calibration. State-of-the-art weak-lensing analyses infer photometric redshift, photo-$z$, distributions by matching the properties of the source galaxies (in this case, fluxes and colors) to representative calibration samples \citep[e.g.,][]{masters_2015,buchs_2019, myles_2021, kids_1000_redshifts, yin_2025,  kids_legacy_redshifts}. We argue that this framework can be extended to intrinsic alignments. Rather than assigning a single broad prior to an entire source sample,  direct IA measurements can serve as an empirical calibration library, with their relevance determined by the overlap between the astrophysical properties of the measured populations and those of the weak-lensing sources. Importantly, this approach also identifies populations for which existing measurements provide insufficient calibration, allowing uncertainty to be retained rather than replaced with extrapolation.

Such a framework also offers a route to optimizing source selection by constructing galaxy populations with intrinsically weak alignments \citep[e.g.,][]{hirata_2007,krause_2015,blueshear}. In particular,  star-forming galaxies have generally been found consistent with negligible large-scale IA \citep[e.g.,][]{mandlebaum_blue_ia,heymans_2013_wl_ia_blue_red, subaru_blue_ia, johnston_2019,samuroff_2023,siegel_2025,paus_direct} (though some spin-axis alignment has been observed in very low-redshift samples at small-scales \citep{small_blue_ia_i,small_blue_ia_ii,small_scale_blue_iii} and spirals are theorized to have different alignment mechanisms than their elliptical counterparts, further complicating their detection \citep{tidal_spiral_ia}. It has been demonstrated that such `Blue Shear' samples can recover competitive cosmological constraints while reducing sensitivity to IA \citep{blueshear}.  However, even these selections require robust calibration of potential residual IA \citep[e.g.,][]{blueshear_bigwood}, particularly where photometric colors do not uniquely determine IA galaxy properties. A demographic approach can therefore inform both the IA priors assigned to a sample and the selection of source galaxies with well-understood alignment properties.


In this paper, we introduce an empirical framework for tailoring IA priors to the galaxy populations used in weak-lensing cosmology. The framework combines existing direct IA measurements -- with their properties compiled into a common library in Sec.~\ref{sec:library} --  with the inferred physical properties of source galaxies to derive informed, population-dependent priors using the overlap between their respective galaxy populations (Sec.~\ref{sec:methods}). As a proof-of-concept, we demonstrate the framework using selections on the DES Year 3 weak-lensing deep sample in the COSMOS field (Sec.~\ref{sec:results}), including selections on physical properties and photometry that integrate with common redshift inference infrastructure. Having introduced and demonstrated a foundation for astrophysically informed intrinsic alignment inference and offering guidance for future weak-lensing sample design, we discuss applications and ramifications of the tailored method in Sec.~\ref{sec:disc}, and conclude in Sec.~\ref{sec:concl}.


\section{The landscape of direct IA measurements}\label{sec:library}

Direct measurements of intrinsic alignments have established that alignment strength depends on a range of galaxy properties and observational choices, including morphology \citep{kids_bright_color_lum_morph_scale}, photometric band \citep{georgiou_ia_wavelength}, shape measurement method \citep{singh_2016}, satellite fraction \citep{Zhou_2023_desy1_redmapper,georgiou_satellite,johnston_2019,gama_satellite_ia}, halo mass \citep{fortuna_2025,fortuna_2021_halomodel}, and luminosity \citep[e.g.,][]{paus_direct,siegel_2025,hervas_peters_2024_unions,johnston_2019,singh_2015,joachimi_megaz_2011}. These trends are not necessarily independent, but are connected to galaxy environment \citep{halo_alignment_peak_height,environment_ia_sim,maion_lagrange_estimator_peak_height,nbody_halo_IA_high_underdense} and formation history \citep{redshift_evolution_lrgs_flamingo}, which also influence galaxy composition and intrinsic properties \citep[e.g.,][]{environmental_link_to_morphology_2026,quenching_cosmic_noon_2025}. A growing compilation of these measurements is available through the echoIA repository\footnote{\url{https://github.com/echo-IA/IAmeasurementsStore}}. Together, these findings motivate characterizing IA in terms of the astrophysical properties of the galaxies being analyzed, rather than treating the source population as homogeneous.


Fig.~\ref{fig:diagnost} summarizes some of the astrophysical properties of galaxy populations (colored contours) that comprise the current observational landscape and their measured amplitudes of alignment (colorbar). For more detail on any one of these samples, the reader is directed to App.~\ref{app:overlap_dm} (see Table~\ref{tab:direct_meas}). Note that our calibration library is limited in this work by the public availability of the matched shape-redshift catalogs for each measurement, and corresponding inferences on their Spectral Energy Distributions (SEDs) that allow for estimation of stellar mass and specific star-formation rates. For this reason, the populations depicted in Fig.~\ref{fig:diagnost} are archetypal of modern direct measurements, though certainly not comprehensive. We make use of the: DESI Y1 measurements developed in \cite{siegel_2025} -- modified with sample splits optimized for this work -- for Luminous Red Galaxies (LRGs, 4 bins), the Bright Galaxy Survey (BGS, 2 bins), and  Emission Line Galaxies (ELG); PAUS measurements in \cite{paus_direct} for their blue (3 stellar mass bins) and red (3 stellar mass bins) samples; and the eBOSS ELG measurements in \cite{samuroff_2023}.

Several trends are apparent from the works in Fig.~\ref{fig:diagnost}: (1) IA vary systematically with intrinsic color, shown here by a trend in specific star-formation rate, with quiescent galaxies exhibiting substantially stronger alignments than star-forming populations; (2) IA amplitude increases similarly with stellar mass, though only significantly for passive galaxies. In the top panel of Fig.~\ref{fig:diagnost} we showcase the single-power-law fit relating alignment amplitude to luminous red galaxy halo-mass from \cite{fortuna_2025}, converted to stellar mass using the galaxy populations in that work (black points) for direct comparison to the bottom panel.

Different photometric surveys performed on instruments with variable bandpasses, exposure times, and observing conditions naturally probe unique galaxy populations. For cosmic shear analyses, the additional splitting of these samples into tomographic bins further changes the astrophysical character of the galaxies detected. In Fig.~\ref{fig:diagnost}, the gray shaded contours represent the span in properties for a deep photometric galaxy population (discussed further in Sec.~\ref{sec:cosmos}), which covers a notably different parameter-space than existing direct measurements -- dominated by bluer, lower-mass galaxies. A central challenge for empirical IA calibration is therefore to extend direct measurements to the galaxy populations that remain poorly represented, and to determine which existing measurements are most applicable to photometric source galaxies.

\section{Method: From Galaxy Populations to Tailored IA}\label{sec:methods}
In this section, we introduce a framework for constructing IA priors  tailored to the astrophysical demographics of photometric weak-lensing source samples. The approach uses direct IA measurements of galaxies with similar properties to inform the calibrated component of the prior, while explicitly accounting for the remaining population through an assumed prior. 
We outline the mathematical prescription in Sec.~\ref{sec:tailored}, discuss the practical implementation and possible extensions in Sec.~\ref{sec:practical}, and consider alternative approaches in Appendix~\ref{sec:alternate}.

\subsection{Tailoring NLA priors on alignment to galaxy demographics} \label{sec:tailored}

\begin{figure*}
    \centering \includegraphics[width=0.99\linewidth]{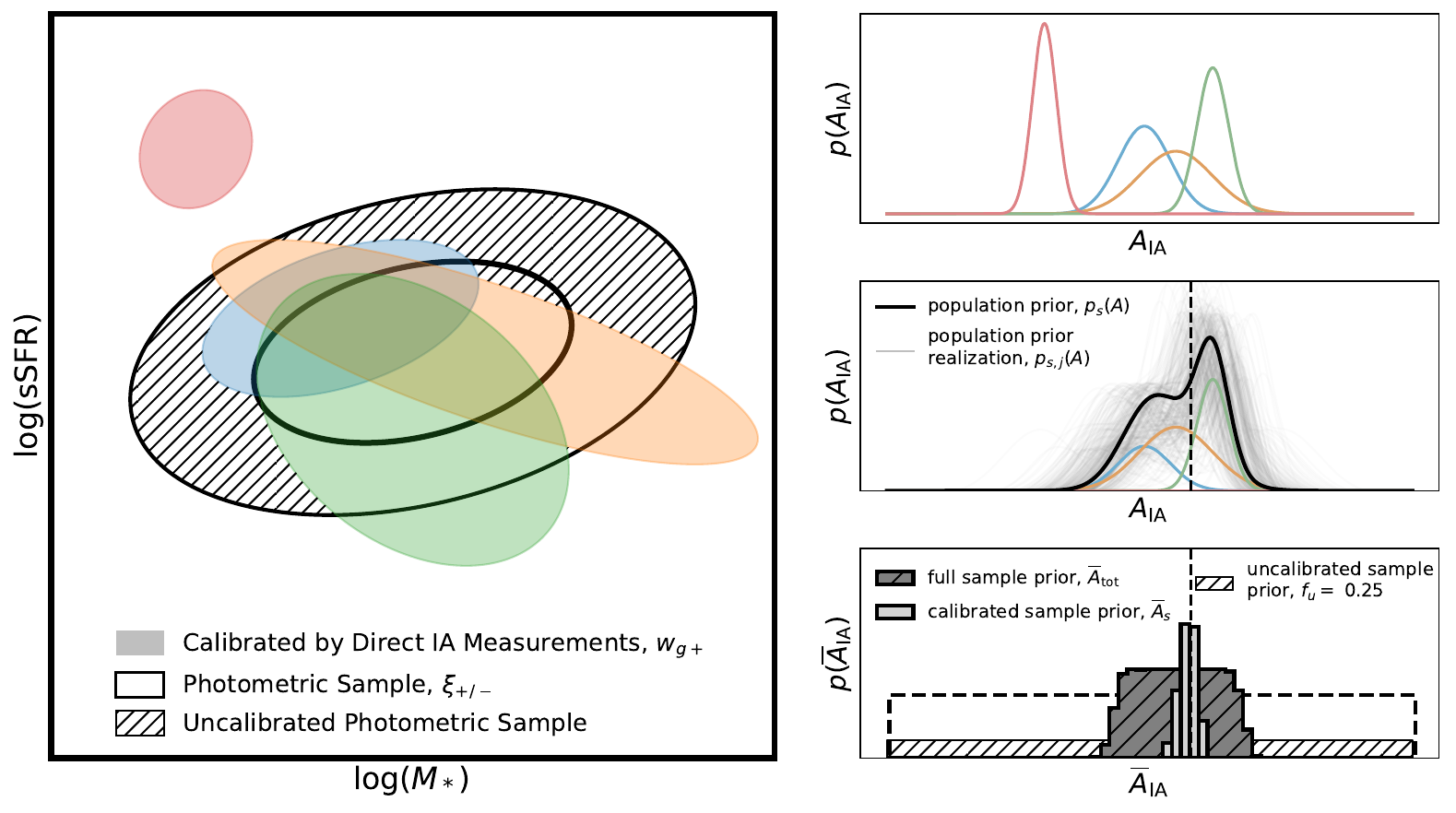}
    \caption{A simple schematic showing the process of leveraging overlap in the galaxy population properties between photometric samples and those from direct measurements of alignment.\\ (\textbf{Left}) An analog of the bottom panel of Fig.~\ref{fig:diagnost}, direct measurement samples (shaded) with measured amplitudes of alignment are plotted as a function of star-formation and stellar mass, alongside the desired photometric sample (black), with some fraction of that sample remaining uncalibrated (hatched).\\(\textbf{Right top}) The posterior for each direct measurement on the amplitude of alignment $A_{\rm{IA}}$ for their respective galaxy populations.\\(\textbf{Right center}) The corresponding distribution of inferred amplitudes for the overlapping population in the photometric sample, $p_s(A)$, taking the top panel amplitudes and reweighting them according to their respective overlap in physical properties. Marginalizing over the contribution of uncertainty in direct measurement amplitude (gray), we create different realizations sampled from their posteriors.\\(\textbf{Right bottom})  The final alignment prior for two-point statistics for the entire photometric population (hatched, dark histogram) is the calibrated-fraction-weighted combination of that derived for the calibrated sample (light histogram, from the means of the many realizations in the previous panel) and that of the uncalibrated sample (uniform prior, hatched, with dashed representing $f_u=1$). Relative normalizations are arbitrarily chosen for readability.}
    \label{fig:example}
\end{figure*}

Direct IA measurements infer the average alignment of a galaxy population from correlations between galaxy positions and shapes. 
In a simplified description, the effective alignment amplitude of a population can be expressed as a weighted average of the amplitudes of its constituent galaxies,
\begin{equation}\label{eqn:ia_mean}
    \langle A_{\mathrm{IA}}\rangle = \frac{\sum_{i\in \rm{g}} w_iA_i}{\sum_{i\in {\rm g}}w_i},
\end{equation}
where $A_i$ denotes the effective NLA alignment amplitude associated with galaxy $i\in \rm{g}$, in the overall population, \citep{bridle_king_2007} and $w_i$ represents its contribution to the measurement; while we never measure $A_i$ for each galaxy independently, choice of $w_i$ would be survey-dependent and is often a function of shape noise. 

Each direct IA measurement population shown in Fig.~\ref{fig:diagnost} has been used to measure the projected position--shape correlation function, $w_{{\rm g}+}$, and to constrain an IA model. Here, we use measurements obtained with the Nonlinear Alignment model \citep[NLA; ][]{bridle_king_2007,hirata_seljak_2004}, as they are the most readily available and provide a posterior for a single amplitude of alignment, $A_{\rm{IA}}$. 

Our prescription has three steps. First, we quantify the demographic overlap between the target source population and the available direct IA calibrators, defining an effective uncalibrated fraction, $f_{\rm u}$. Second, we combine the direct measurements to construct an empirically informed distribution of IA amplitudes, $p_{\rm s}(A)$, representing the calibrated contribution to a target photometric sample. Finally, we propagate the measurement uncertainties and account for incomplete calibration to obtain a prior on the effective mean IA amplitude of the full target population, $p(\overline{A}_{\rm tot})$.

\subsubsection{Quantifying demographic overlap}
When performing inference from direct measurements, we would like to characterize how astrophysically represented one sample is by another -- requiring us to isolate any demographic properties correlated with IA. For this preliminary work, we quantify the similarity between galaxy populations using their joint distributions of stellar mass and specific star-formation rate, with $x=\log(M_*)$ and $y=\log(\mathrm{sSFR})$. For two populations $i$ and $j$, we define the demographic overlap coefficient, $\mathcal{O}_{ij}$, as
\begin{equation}\label{eqn:overlap}
    \mathcal{O}_{ij}(p_i,p_j) = \iint \mathrm{min}\left[p_i(x,y),p_j(x,y)\right] \mathrm{d}x \mathrm{d}y,
\end{equation}
where $p_i(x,y)$ and $p_j(x,y)$ are their normalized two-dimensional distributions. $\mathcal{O}_{ij}$ equals unity for identical distributions (a perfect calibration) and decreases as their demographics differ. Examples of the 68\%-regions of $p(x,y)$ distributions for both photometric and direct measurement populations are shown in Fig.~\ref{fig:diagnost}.

We use this overlap to quantify the extent to which the photometric source population is represented by the available direct IA measurements.
Here, we refer to each galaxy sample with a direct IA measurement as a \textit{calibrator population} and to the photometric source sample for which we wish to construct an IA prior as the \textit{target population}. For a single calibrator population, we define an effective uncalibrated fraction as $f_{\rm u}\equiv1-\mathcal{O}_{ij}$. To account for multiple calibrator populations,  $p_1,p_2,p_3,...\in p_c$, we take the maximum of their normalized distributions at each point in astrophysical parameter space, giving
\begin{equation}
    f_{\rm u}(p_{\rm{phot}},p_{\rm c}) = 1 -\iint \mathrm{min}\left[p_{\rm{phot}}(x,y),\mathrm{max}\left(p_{\rm c}(x,y)\right)\right] \mathrm{d}x\mathrm{d}y.
\end{equation}
Here $p_{\rm phot}$ is the normalized distribution of the target population and $p_{\rm c}$ denotes the set of distributions for the individual calibrator populations. We refer to $f_{\rm u}$ as an effective uncalibrated fraction because it captures differences in relative abundances as well as incomplete coverage of astrophysical parameter space, rather than strictly counting galaxies outside the calibrator populations. This distinction matters because direct IA measurements constrain the mean alignment of a galaxy population: even when a calibrator spans the full range of properties occupied by the target population, differences in the relative abundances of galaxies can affect the applicability of its measured mean alignment. Our overlap metric therefore favors measurements of populations that closely match the target demographics, motivating finer subdivisions of existing direct measurements in larger samples. A schematic illustration is shown in the left panel of Fig.~\ref{fig:example}.

\subsubsection{Combining direct IA measurements}
For each calibrator population $i$, we approximate the posterior on its measured NLA amplitude as a Gaussian, $\mathcal{N}(A_i,\sigma_i^2)$, with mean $A_i$ and variance $\sigma_i^2$. For the segment of the target population that overlaps direct measurements, ${\rm s}$, we construct an effective distribution of IA amplitudes, $p_{\rm s}(A)$, by linearly combining these measurements. 
Each measurement is weighted by its demographic overlap with the target population, $\mathcal{O}_{is}$, and by its inverse variance, so that more precise measurements are strongly favored over others with comparable demographic overlap:
\begin{equation}\label{eqn:combination}
    p_{\rm s}(A) = \frac{\sum_{i\in c} (\mathcal{O}_{i {\rm s}}/\sigma_i^2)\ \mathcal{N}(A_i,\sigma_i^2)}{\sum_{i\in c} (\mathcal{O}_{i{\rm s}}/\sigma_i^2)}.
\end{equation}
The above equation assumes the overlap between individual direct measurement samples is small and neglects any correlation. When calibrator samples do overlap or their measurements have correlated uncertainties, we can account for these correlations using a covariance matrix $C_{ij}$, replacing the inverse-variance weights with
\begin{align}
    p_{\rm s}(A) &= \frac{\sum_{i\in c} \mathcal{O}_{i{\rm s}}w_i\ \mathcal{N}(A_i,\sigma_i^2)}{\sum_{i\in c} \mathcal{O}_{i{\rm s}}w_i},\ \,\, \mathrm{where} \,\,\,\,
    w_i = \sum_{j\in c} C_{ij}^{-1}\, .
\end{align}
Further details of the covariance estimation are provided in App.~\ref{app:overlap_dm}. This weighted combination step is depicted in the center right panel of Fig.~\ref{fig:example}. We note that inverse-variance weights are likely suboptimal in this scenario, as each $A_i$ is not an independent measure of the same quantity, but rather built from a unique population of galaxies with some shared features. Examples can be constructed that make this weight definition a potential source of bias (e.g.,~an exquisite measurement of the $A_i$ of a small, non-representative subsample where $\sigma_i\approx0,\ \mathcal{O}_{is}\ll1$, or an extremely uncertain measurement of a well-overlapping subset). As such conditions do not exist among our current library of IA direct measurements and their interpretation would require a strong prior on the parameterization of alignment, 
we defer a more complete treatment to future work. 

Although we refer to $p_s(A)$ as a `population-level' distribution of IA amplitudes, it should not be interpreted as the true distribution of individual galaxy alignments, but rather the combined distributions of sub-population means. Each direct IA measurement constrains the mean alignment of a population spanning a range of stellar masses and specific star-formation rates, and is subject to measurement uncertainty. The resulting distribution is therefore limited by both the precision of the measurements and how finely their galaxy populations resolve the astrophysical properties of the target sample. 

Our combination (Eqn.~\ref{eqn:combination}) further assumes that the mean alignment measured for each calibrator population, $A_i$, is representative of the galaxies with similar properties in the target population -- which will always be true only if alignment is constant for all galaxies in a given sample, e.g., in the null case. This approximation can be improved either by remeasuring IA for more closely matched galaxy populations to the target sample (our recommendation) or by correcting existing measurements for sample differences using a model for the dependence of alignment on galaxy properties, as discussed in Sec.~\ref{sec:practical} and elaborated upon in App.~\ref{sec:alternate}.

\subsubsection{Constructing the sample-level IA prior}
For cosmic shear cosmological analyses, we seek a prior on the effective mean alignment amplitude of the source sample, rather than the distribution of amplitudes represented by $p_s(A)$. We first compute the mean of the empirically informed distribution,
\begin{equation}\label{eqn:mean_aia}
    \overline{A}_{\rm s} = \frac{\int_a^b Ap_{\rm s}(A) \, \mathrm{d}A}{\int_a^b p_{\rm s}(A) \, \mathrm{d}A} \, ,
\end{equation}
where $a$ and $b$ define the integration bounds. We adopt $(a,b)=(-10,10)$ to allow for contributions from strongly aligned populations.

To propagate uncertainties in the direct IA measurements, we repeat this calculation for $n$ realizations, drawing an amplitude $A_{i,j}$ from the Gaussian posterior of each calibrator population $i$ in realization $j$ and using $\mathcal{N}(A_{i,j}, \sigma_i^2)$ as input to Eqn.~\ref{eqn:combination}. These draws produce a family of distributions $p_{s,j}(A)$ and corresponding means $\overline{A}_{s,j}$, whose distribution defines the prior on the effective mean amplitude of the calibrated component of the target sample. This procedure propagates the uncertainty in the measured amplitudes, rather than treating the full width of $p_s(A)$ as uncertainty in the sample mean. 
A demonstration of this step can be seen in the center right panel of Fig.~\ref{fig:example}, showcasing the realizations of the population-level distribution (gray), with lower right panel showing the resulting distribution of means (light gray, histogram). 

We next account for the component of the target population that is not represented by the direct IA measurements. As the mean of a sample can be written in terms of the means of its two sub-populations, we combine the calibrated mean amplitude with an unknown amplitude for this component using
\begin{equation} \label{eqn:combine_frac}
    \overline{A}_{\rm{tot}} = (1-f_{\rm u})\overline{A}_{\rm s}+f_{\rm u}\overline{A}_{\rm{uncal}} \,.
\end{equation}
Here, $f_{\rm u}$ is the effective uncalibrated fraction, rather than a literal fraction of galaxies assigned to an uncalibrated subset. Equation~\ref{eqn:combine_frac} therefore provides an effective mixing prescription, motivated by the abundance-weighted mean of two populations, in which demographic mismatch determines the relative contribution of the empirically informed and unconstrained amplitudes.

Since the mean amplitude of the uncalibrated component, $\overline{A}_{\rm{uncal}}$ is unknown, we assign it a wide, flat uniform prior, $\overline{A}_{\rm{uncal}}\sim U(-x,x)$, where $x=5$ for our fiducial choice (though we also explore $x=3$).  For each realization $j$, we draw an uncalibrated amplitude independently and compute the combined prior on the effective mean alignment amplitude for the full photometric sample, $\overline{A}_{\rm{tot},j}$, as:
\begin{equation}\label{eqn}
    \overline{A}_{\rm{tot},j} = (1-f_{\rm u})\overline{A}_{s,j}+f_{\rm u}\,\overline{A}_{\rm{uncal},j}.
\end{equation}
This combination of the calibrated and uncalibrated priors is depicted as the dark gray, hatched histogram in the bottom right panel of Fig.~\ref{fig:example}. 

Alternatively, if the distribution of calibrated mean amplitudes is well approximated by a Gaussian, $\overline{A}_s\sim\mathcal{N}(\mu_s,\sigma_s^2)$, the final prior can be obtained analytically by convolving this Gaussian with the uniform distribution:
\begin{multline}\label{eqn:analytic}
    p(\overline{A}_{\rm{tot}}) = \frac{1}{4f_{\rm u} x} \left( \rm{erf} \left[\frac{\overline{A}_{\rm{tot}}-(1-f_{\rm u})\mu_s+f_{\rm u} x}{(1-f_{\rm u})\sigma_s\sqrt{2}}\right]\right. \\
    -\left.\ \rm{erf}\left[\frac{\overline{A}_{\rm{tot}}-(1-f_{\rm u})\mu_s-{f_{\rm u}} x}{(1-{f_{\rm u}})\sigma_s\sqrt{2}}\right]\right). 
\end{multline}
For an intuition on how the magnitude of the uncalibrated fraction mediates the final shape of the distribution and the agreement of the analytic form with the numerical sampling, see App.~\ref{app:uncal_dists}.


\subsection{Practical considerations and future extensions}\label{sec:practical}
\textbf{Obtaining astrophysical properties for photometric galaxies.} For the galaxy populations used in direct IA measurements -- often observed with spectroscopy or medium- and narrow-band photometry to obtain high-quality redshift estimates -- stellar masses and specific star-formation rates are well-constrained (see Table~\ref{tab:direct_meas} for more information). For photometric source populations, these properties are more challenging to infer. In this work, we therefore focus on the COSMOS deep field, where extensive medium- and narrow-band photometry, together with optical and near-infrared observations, enables detailed characterization of galaxy properties \citep{cosmos_2025}. We discuss the extension to wide-field shear catalogs with more limited photometric information in App.~\ref{app:wide} (see also Sec.~\ref{sec:som}), where we leverage existing survey infrastructure. 

\textbf{Methodological differences in astrophysical-property inference: }Inference for galaxy properties from spectroscopic or photometric data can be subject to very different assumptions that can contribute to offsets on those parameters between surveys -- with deviations on the order of 0.2 dex in stellar mass in particular even when fitting the same data \citep[e.g., ][]{bellstedt_sed}. These deviations likely exist between samples in our compiled library of direct measurements, (see Jeffrey et al., in prep, \citealt{fastspecfit}), but are not yet accounted for in our proof-of-concept demonstrations. 
Because these offsets affect the inferred demographic overlap, future applications should ideally derive galaxy properties consistently across the calibrator and photometric populations. Alternatively, uncertainties and biases in the relative property estimates could be propagated by allowing shifts in the distributions $p_i(\log\mathrm{sSFR},\log M_*)$ when computing the overlap coefficients.

\textbf{Extension to higher dimensions: }Existing direct IA measurements have not established any significant trend in IA strength as a function of galaxy redshift that is not accounted for by evolution in other factors like color and mass \citep[e.g.,][]{joachimi_megaz_2011,mandlebaum_blue_ia, johnston_2019,siegel_2025, paus_direct}, though work in simulations has suggested this evolution could be present \citep[e.g.,][]{redshift_evolution_lrgs_flamingo, nbody_halo_IA_high_underdense}. We therefore use stellar mass and specific star-formation rate as the two demographic axes in this proof-of-concept. The framework can be extended to include redshift or other relevant properties, such as central or satellite probability, provided these can be reliably inferred for the photometric source population. The major cost to incurring additional dimensions is the reduced overlap with the limited, direct measurement populations across joint, high-dimensional space. Establishing which additional properties are necessary to characterize alignment will therefore help identify where new direct or indirect  \citep[e.g.,][]{self_calibration_i,selfcalibration_pederson} IA measurements would be most valuable. 

\textbf{Computing the 2D distribution $p_i$ and weighting:} We characterize each galaxy population by its normalized joint distribution in stellar mass and specific star-formation rate,  $p_i(\mathrm{log\ sSFR},\mathrm{log}\ M_*)$. 
For a well-sampled selection of galaxies, we construct a two-dimensional histogram on a $200\times200$ grid. When the distribution tails are sparsely populated, we instead use a Gaussian kernel density estimate\footnote{This is implemented using \texttt{scipy.stats.gaussian\_kde} \citep{scipy,multivariate_kde}.}  to smooth the distribution; this produces results that are insensitive to binning choice.  
 We choose the limits $6.5 < \mathrm{log}(M_*) < 12$ and $-14 < \mathrm{log}(\mathrm{sSFR}) < -7$, assigning galaxies outside these ranges to an additional row or column of `pileup'. The lower sSFR bound accommodates the substantial uncertainty in star-formation rates for quenched galaxies. 
 
 Ideally, these distributions would incorporate the same galaxy weighting used in a cosmic shear analysis. Furthermore, we recommend remeasuring the direct IA amplitudes after reweighting each calibrator population to match the astrophysical distribution of the target galaxies with which it overlaps. This improves our ability to infer unbiased alignment by reducing our method's reliance on the assumption that a calibrator's mean alignment is representative of its overlapping population. Recomputing the reweighted sample's direct measurement is the preferred choice, rather than requiring a strongly parameterized model of IA dependence on galaxy properties to account for alignment gradients (see App.~\ref{sec:alternate}). We leave this refinement and re-measurement to future work, but note our recovery of unbiased results with the simulated samples in App.~\ref{sec:alternate} when this astrophysical weighting is employed.

\textbf{Choice of alignment model and use of direct measurement posteriors: } Although we demonstrate the method using the NLA model, the general approach could be extended to other IA models with additional parameters \cite[e.g.,][]{eft, blazek_2019,halomodel}. Such extensions would need to preserve correlations between parameters rather than constructing independent priors where significant degeneracies exist. 
The ideal implementation of this method would similarly have the IA model posteriors on the direct measurements within the same shear catalog being calibrated, and draw from those posteriors rather than the Gaussian approximation in this work. 
For this proof-of-concept, we also make a simplifying choice for low-significance measurements: when a direct IA amplitude is consistent with zero within $2\sigma$, we re-center its Gaussian distribution at zero while retaining the reported uncertainty, as seen in Fig.~\ref{fig:diagnost}. 
This is an explicit assumption that these populations have negligible mean alignment and diminishes the impact of scatter within under-confident measurements. The choice affects the resulting calibrated priors and should be revisited in future applications using the full measured posteriors. In App.~\ref{sec:alternate}, we demonstrate for this work that this assumption, for a single choice of analytic IA model, does not shift our result for any sample beyond their 68\% regions.

\textbf{Per-tomographic-bin priors and intra-bin variation: }A natural application of this framework is to construct separate IA priors for the source populations within each tomographic bin of a cosmic shear analysis. However, broad tomographic bins can contain galaxies with substantially different astrophysical properties and IA amplitudes. Subdividing these populations, where the photometric information permits, could help characterize this variation, although the resolution is ultimately limited by the available direct measurements and degeneracies in the inferred galaxy properties.

An additional limitation arises because the `GI' (the correlation between an intrinsically-aligned shape and a background gravitationally-lensed shape)  and `II' (intrinsic-intrinsic correlation) contributions to cosmic shear depend differently on alignment amplitude. Under the NLA model, GI is linear in the amplitude, whereas II is quadratic \citep[e.g.,][]{lamman_review}. Consequently, the mean amplitude inferred here does not generally capture the full II contribution when alignment varies within a source population. Higher moments of the inferred amplitude distribution may provide useful information, but a consistent treatment would also need to account for correlations between the alignments of different galaxy populations. We leave such extensions to future work.

\textbf{Choice of prior for the uncalibrated galaxy populations: }We adopt a wide, flat prior, $\mathcal{U}(-5,5)$, for the mean alignment of the uncalibrated component, broadly encompassing the range of NLA and analogous tidal-alignment amplitude priors used in previous cosmic shear analyses \citep[e.g.,][]{amon_2021,secco_2021,kids1000_cosmic_shear,hsc_fourier,hsc_real}, though narrower, asymmetric priors have been used more recently, e.g., $A_{\rm{IA}}\sim\mathcal{U}(-1,3)$ for DES Y6 \citep{desy6_cosmic_shear}.  In principle, the prior could be informed by the astrophysical properties of the uncalibrated population, such as adopting a narrower distribution for star-forming galaxies, e.g., a Gaussian centered at zero for bluer samples \citep[][]{blueshear_bigwood} or with less extreme bounds for lower stellar masses.  
However, such assumptions risk excluding significant alignment in poorly measured populations, including low-mass satellites. We therefore retain $\mathcal{U}(-5,5)$ as our fiducial choice and explore $\mathcal{U}(-3,3)$ to illustrate the sensitivity of the final prior to this assumption.



\section{Proof-of-concept with DES}\label{sec:results}

In this section, we demonstrate the Tailored-IA method by constructing emperical IA priors for photometric selections of galaxies within the COSMOS deep field. First, we describe the galaxy sample and its inferred properties (Sec.~\ref{sec:cosmos}), before deriving priors for quenched and star-forming galaxy selections (Sec.~\ref{sec:bluered}). We then apply the method for many narrow color-selections within the Self-Organizing Map (SOM) used for photometric-redshift calibration (Sec.~\ref{sec:som}), including an illustrative selection in redshift and stellar mass designed to have strong overlap with existing direct IA measurements (Sec.~\ref{sec:tomo}). The resulting IA model priors are summarized in Table~\ref{tab:prior_outputs}.

For completeness we explore results for the full, uncut sample in App.~\ref{app:full}, and for a comparison of the different photometric selections, see App.~\ref{app:phot_properties}. Comparisons to more analytic methods of producing an alignment prior are relegated to App.~\ref{sec:alternate}. 

\subsection{Dark Energy Survey: COSMOS}\label{sec:cosmos}

We demonstrate the tailored-IA method using galaxies in the COSMOS deep field, where the combination of Dark Energy Survey (DES; \citealt{DES}) photometry and extensive ancillary observations enables detailed characterization of the astrophysical properties relevant to IA. We use \textit{ugriz} observations from the DES COSMOS deep-drilling field \citep{des_deep_field_y3}, augmented by additional near-infrared photometry in \textit{JHK$_s$}, taken on the Visible and Infrared Survey Telescope for Astronomy \citep{vista}.

To obtain precise estimates of stellar mass and star-formation rate, we use the COSMOS2025 catalog\footnote{\url{https://cosmos2025.iap.fr/catalog.html}} \citep{cosmos_2025}, which uses additional medium- and narrow-band observations to inform spectral energy distribution fits performed with the \textsc{cigale} algorithm \citep{cigale_boquien}. The star formation rate is defined as 
\begin{equation}
    \mathrm{log(sSFR)} = \mathrm{log}(\texttt{cigale\_sfr\_inst}) - \mathrm{log}(\texttt{cigale\_mass})
\end{equation}
and galaxy properties are plotted as the shaded gray contour in Fig.~\ref{fig:diagnost}.
After cross-matching to the DES deep-field catalog, we have 114,559 galaxies in this sample with well-measured stellar masses and star-formation rates. We apply the recommended quality criteria\footnote{selections: $\texttt{cat['type'] == 0, }
    \texttt{cat['mag\_model\_f444w'] < 30, } \\
    \texttt{cat['warn\_flag'] == 0, }
    \texttt{cat['flag\_star\_hsc'] == 0}$} that restricts us to galaxies with reasonable signal-to-noise outside the star mask:




\begin{figure*}
    \centering\includegraphics[width=\linewidth]{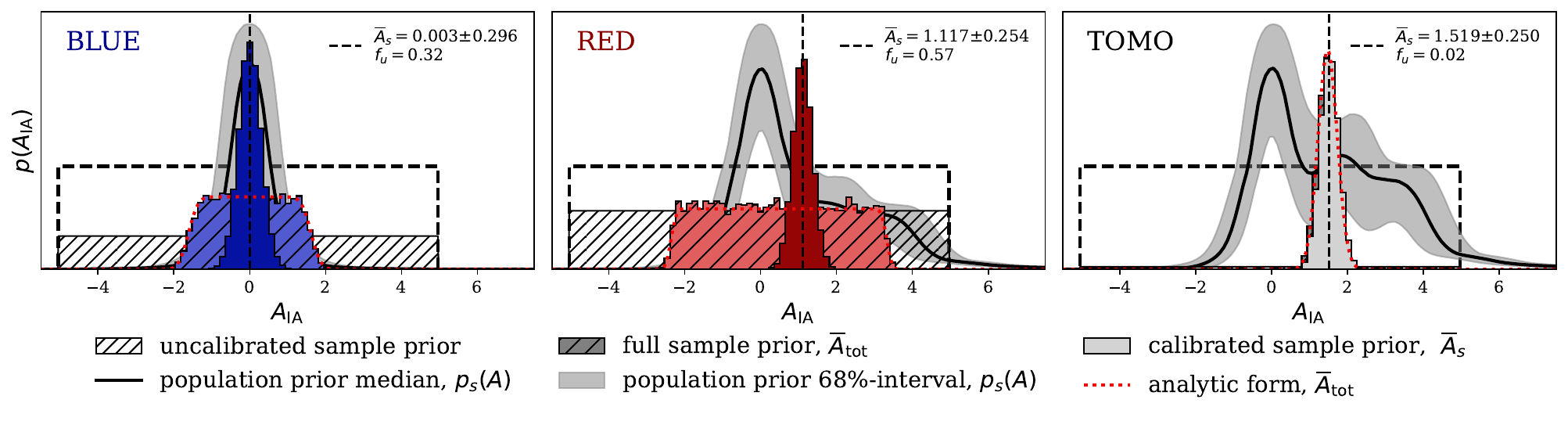}
    \caption{For the \textsc{blue} (left), \textsc{red} (center),  and \textsc{tomo} (right) selections, we plot the population-level distribution of alignments for their measurement-overlapping segments ($p_s(A)$, line) along with the 68\%-interval of their bootstrapped realizations (gray shaded). For these realizations we plot their means (dark histogram) that represent the prior on alignment for the calibrated sample, $\overline{A}_s$. When combined with the flat prior set on the uncalibrated component of the population (hatched, visually weighted by $f_u$ within dashed region), we also display the final priors for the whole sample (light histogram, hatched), $\overline{A}_{\mathrm{tot}}$. Note that relative normalizations are chosen for readability and are not physical. See results summarized in Table~\ref{tab:prior_outputs}. The \textsc{tomo} selection is chosen to have a high calibration fraction, and is not representative of any practical weak-lensing source population.}
    \label{fig:per_gal_prior}
\end{figure*}

\subsection{For \textsc{red} and \textsc{blue} deep samples}\label{sec:bluered}

Because passive, quenched (often `early-type' elliptical) galaxies exhibit stronger IA than actively star-forming galaxies (often `late-type' spirals) \citep[e.g.,][]{chisari_review}, separating weak lensing samples by galaxy type has been used both as a robustness test \citep[e.g.,][]{heymans_2013_wl_ia_blue_red,desy1_color_split,kids_legacy_consistency} and as an IA mitigation strategy \citep[e.g.,][]{blueshear,popcosmos_properties}. Here, we use the well-defined astrophysical properties available in the COSMOS deep field and defined in Sec.~\ref{sec:cosmos} to construct separate IA priors for quenched \textsc{red} ($\mathrm{log}(\mathrm{sSFR}) \leq -10$) and star-forming \textsc{blue} ($\mathrm{log}(\mathrm{sSFR}) > -10$) populations. 
Note that the delimiter is drawn as a dashed red line in Fig.~\ref{fig:diagnost}. The \textsc{full} sample of galaxies is dominated by \textsc{blue} galaxies ($86$k) with fewer \textsc{red} galaxies ($24$k). Applying the framework of Sec.~\ref{sec:methods}, we depict the inference for demographic IA amplitude in Fig.~\ref{fig:per_gal_prior}. 

For the \textsc{blue} sample (left panel), the inferred mean alignment amplitude of the calibrated population is $\overline{A}_s =0.00\pm0.30$, consistent with no intrinsic alignment (with $\pm$ indicating a 68\% interval, and not a Gaussian posterior). The solid line represents the per-galaxy alignment distribution $p_s(A)$ with the shaded region indicating the 68\%-interval across $n = 10,000$ realizations that marginalize over measurement uncertainty. The distribution of their means (dark blue histogram), defines the prior on the calibrated population. However, an uncalibrated fraction $f_u=0.32$ of the sample lies outside effective calibration coverage. Marginalizing over this uncalibrated component using the broad uniform prior (hatched) yields a substantially wider prior for the full blue population, $\overline{A}_{\mathrm{tot,\ }\textsc{blue}}=0.01\pm1.10$ (light blue histogram, hatched). Note, restricting the uncalibrated prior to $\mathcal{U}(-3,3)$ narrows the final constraint by  $\sim 1.7$ to $\overline{A}_{\mathrm{tot,\ }\textsc{blue}} = 0.00\pm0.66$, illustrating its sensitivity to assumptions about galaxies without direct IA calibration.

For the \textsc{red} sample (middle panel), the calibrated population exhibits a significant positive alignment amplitude, $\overline{A}_s =1.12\pm0.25$. 
Despite this relatively precise measurement, the uncalibrated fraction is substantially larger, $f_u=0.57$, primarily because low-mass red galaxies lack direct IA calibrators. Consequently, the final prior for the full red population is considerably broader, $A_{\mathrm{tot,red}}=0.47\pm1.94$. This illustrates a central feature of the tailored-prior approach: the precision of the final IA prior depends not only on the precision of existing direct measurements, but also on how well those measurements cover the astrophysical properties of the source population.

We note that the \textsc{red} sample has substantially poorer calibration coverage than the \textsc{blue} sample in the $M_*$–sSFR plane, owing to poor overlap at low stellar mass -- though direct measurements are typically more precise than blue samples for the strongly-aligning amplitudes at higher stellar masses. However, these results do not account for redshift dependence in IA beyond that captured by galaxy evolution in our chosen demographics. 

If redshift evolution in IA measurements was eventually observed and this axis was added to our inference, the calibrated fraction of the \textsc{blue} sample would drop significantly, particularly for high ($z\gtrsim1.3$) and low ($z\lesssim0.3$) populations. The former would be especially relevant for the highest redshift bins in cosmic shear surveys. Finally, this demonstration does not account for gradients in alignment strength within individual direct measurement samples; we explore their potential impact in App.~\ref{sec:alternate}. 

\begin{figure*}
    \centering \includegraphics[width=\linewidth]{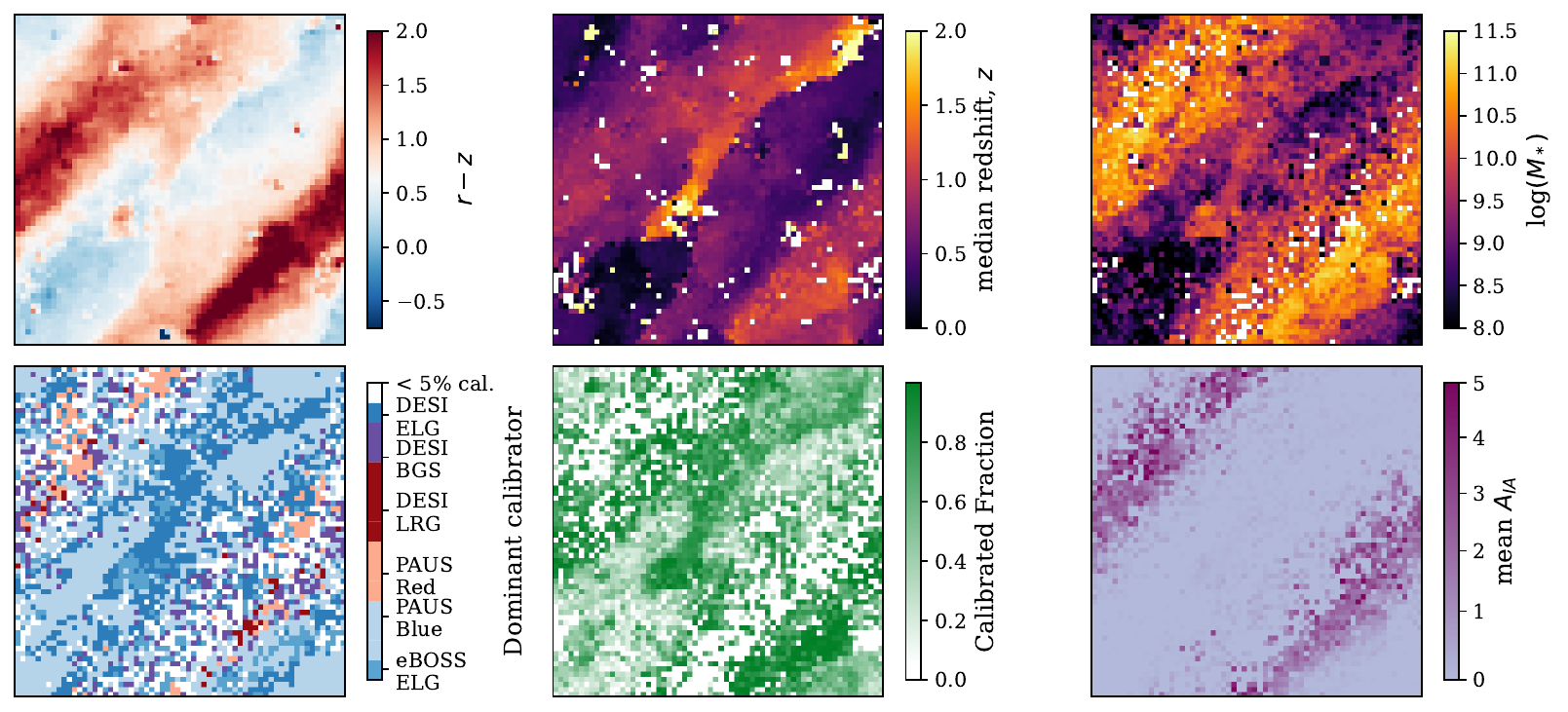}
    \caption{The DES Y3 Self-Organizing Map (SOM) for the deep field derived in \protect\cite{myles_2021}, that compresses the \textit{ugrizJHK$_s$} color-space to two axes, each cell of which comprises a selection on optical and near-infrared photometry. Pictured in this map is the observed $r-z$ color (top left), the median redshift of the calibrator samples in each cell used for photometric redshift inference (top center), the median stellar mass from \protect\cite{cosmos_2025} (top right) and the mean inferred intrinsic alignment amplitude from this work (bottom right).The calibrated fraction (bottom center, 1-$f_u$) and dominant calibrating sample (bottom left) are similarly plotted in the deep SOM. Cells with less than 5\% of their sample overlapping direct measurements are excluded from the bottom left panel (white).}
    \label{fig:allsom}
\end{figure*}

\subsection{For narrow photometric selections: SOM case study}\label{sec:som}
The broad selections detailed in the previous section are performed on inferred properties from high-resolution, many-band photometry, presenting a challenge to translate our inference to weak-lensing surveys that observe flux in only a few bands. The more limited color information in wide-field shear surveys has historically also been an obstacle to inference of photometric redshifts for those populations, which are commonly subject to color-redshift degeneracies \citep[e.g.][]{ newman&gruen}. The discretization schema of color-space and the high-resolution deep-field observations necessary for modern redshift calibration are a very natural scaffolding for any demographic treatment of intrinsic alignment, allowing us to re-weight well-understood galaxies according to their photometric abundances.

Motivated by this, we demonstrate the efficacy of the tailored-IA prior approach in the same deep-field photometric space used for redshift inference in DES Year 3 and Year 6 \citep{myles_2021,yin_2025}. These analyses use Self-Organizing Maps (SOMs; \citealt{Kohonen1982,som_og,kohonen_som_book}) to discretize high-dimensional color space into `cells' that contain galaxies with similar photometric properties via unsupervised machine learning \citep[SOMPZ;][]{buchs_2019,sanchez_2020,myles_2021,campos_2024,yin_2025}. The transfer function between deep- and wide-field observations for this survey was characterized using synthetic-source-injection with \textsc{balrog} \citep{balrog}. Here, we demonstrate the IA inference within only the deep-field population, leaving the propagation to wide-field source galaxies in App.~\ref{app:wide} and future work. 

Fig.~\ref{fig:allsom} shows the `Deep SOM', constructed from $ugrizJHK_s$ photometry, with each cell representing a population of galaxies with similar observed colors \citep{myles_2021}. The upper panels show the median $r-z$ color (left), median calibrator redshift (center), and the calibrator median stellar mass (right) -- which is the mean of a relatively broad distribution per-cell. The lower panels show the direct IA measurement with the largest overlap coefficient $\mathcal{O}$ for each cell (left), the calibrated fraction, $1-f_u$ (middle), and the inferred mean IA amplitude, $\overline{A}_s$ (right), using the method proposed in this work. Although the SOM x- and y- coordinates have no direct physical meaning, the maps reveal how the available IA information varies across photometrically selected galaxy populations, with visible clustering of cells that share demographic features. 

 As expected, bluer populations are predominantly calibrated by emission-line galaxy measurements (DESI and eBOSS ELGs, PAUS Blue), whereas redder populations are better covered by DESI BGS and LRG measurements and PAUS Red. More importantly, the calibrated fraction varies substantially across the SOM, identifying populations for which existing direct IA measurements provide strong constraints and others where significant gaps remain. This map is a useful output of the framework: it identifies where additional direct measurements would be most valuable and where source selections could exclude galaxies with poorly-informed IA. As the inference is performed within the same photometric framework used for redshift calibration, it also provides a practical route toward incorporating demographic IA information into future weak-lensing analyses.
 

\subsubsection{A tailored tomographic selection}\label{sec:tomo}
An important application of the tailored-IA approach is to inform source sample selection. The SOM provides a way to identify populations with both well-characterized redshift distributions and strong overlap with existing direct IA measurements. Selecting such populations could yield tighter, more robust IA priors, which can be weighed against the loss of statistical power from excluding galaxies to optimize the cosmological constraints of a weak-lensing sample.

To illustrate this possibility, we construct a pseudo-tomographic sample by selecting a combination of cells in the deep SOM that occupy a well-calibrated region in the log(sSFR)-log($M_*$) plane and have relatively narrow redshift distributions. The conditions set on each cell $c$ for this selection include
\begin{equation}
    \begin{gathered}
        0.6\ <\ \ \langle{z}\rangle_{\rm c}\ <\ 1.5 \, ,\\
        \sigma_{z,\rm{c}}\ <\ 0.1 \, ,\\
        \langle{\mathrm{log}(M_*)}\rangle_{\rm c}\ >\ 9.5 \, ,\\
        -14\ <\ \langle{\mathrm{log(sSFR)}}\rangle_{\rm c}\ <\ -10.3 \, . 
    \end{gathered}
\end{equation}
Here, $\langle{z}\rangle_{\rm c}$ and $\sigma_{z,{\rm c}}$ denote the mean and standard deviation of redshift within a cell and $\langle{\mathrm{log}(M_*)}\rangle_{\rm c}$ and $\langle{\mathrm{log(sSFR)}}\rangle_{\rm c}$ describe its mean stellar mass and specific-star-formation-rate. These illustrative bounds select 376 of the 4096 SOM cells, containing approximately 1,700 galaxies to demonstrate how a population with coherent redshift and astrophysical properties can be selected to obtain a well-constrained IA prior. We deliberately select a predominantly passive population to demonstrate the inference of a significant, non-zero IA amplitude, rather than the near-zero alignment expected for a blue population. For a census of the relevant properties of this selection, see App.~\ref{app:phot_properties}. While we have hand-tuned this selection on inferred properties to have a low uncalibrated fraction as a simple demonstration, we note that similar selections could be made on $f_{u,\rm c}$ that populate a less coherent galaxy demographic space but would be equally well constrained by direct measurements.

The resulting IA prior is shown in the right panel of Fig.~\ref{fig:per_gal_prior}. For the calibrated 'tomographic' (\textsc{tomo}) population, we obtain $\overline{A}_s = 1.52\pm 0.25$ (dark gray histogram, the means of the 10,000 realizations). Crucially, the selection has a very small uncalibrated fraction, $f_u=0.02$, such that the final prior (light gray histogram) retains nearly the same precision, $\overline{A}_{\mathrm{tot,\ }\textsc{tomo}} = 1.48\pm0.25$. This is approximately $\sim4.5\times$ narrower than the prior for the \textsc{full} sample and $7.7\times$ narrower than that for the \textsc{red} sample (Table~\ref{tab:prior_outputs}). Although this small, illustrative sample is not intended for cosmological analysis as it lacks statistical power, it demonstrates that a tailored approach to source selection can substantially improve IA prior precision.

\section{Discussion}\label{sec:disc}

Our results demonstrate both the potential and the current limitations of constructing empirical, population-dependent IA priors. Existing direct measurements can provide informative constraints for well-characterized galaxy populations, but their incomplete demographic coverage substantially broadens the priors for typical weak-lensing source samples. We discuss how this framework (\textit{i}) complements developments in IA modeling, (\textit{ii}) how it could be incorporated into cosmological analyses, and (\textit{iii}) where additional observations are needed.

\textbf{(i) Tailored priors complement improved IA models}: The framework proposed is agnostic to the underlying IA prescription. Although demonstrated here for a single IA amplitude parameter, the tailored approach can be combined with the nonlinear alignment model, TATT, halo-based models and effective field theory descriptions. Extending this framework to infer multiple parameters would require additional care beyond this work to account for their shared degeneracies. 
As IA models become more sophisticated, with more free parameters, tailored, population-dependent priors offer a way to retain model flexibility without unduly limiting the precision of cosmological constraints. 

\textbf{(ii) Toward astrophysically informed weak-lensing analyses}:
Modern weak lensing analyses already tailor their priors governing photometric redshift calibration to the properties of the galaxies in each tomographic bin. 
We argue that intrinsic alignments should be treated in an analogous way: rather than assigning broad survey-wide priors, or imposing restrictive assumptions about their evolution between bins, IA priors should reflect the demographic properties of the galaxies being analyzed and be robust to existing gaps in our IA knowledge. While we compare our source samples to direct measurements on two property axes (sSFR, $M_*$) in this work, future analyses are not necessarily limited to these specific demographics.

Existing approaches have begun to incorporate galaxy-type information, although they differ in how they treat uncertainty.
For example, the NLA-M model in KiDS-Legacy assumes negligible alignment for blue galaxies \citep{Wright_2025}, while the UNIONS analysis \citep{unions} incorporates a single direct constraint on blue-galaxy alignments \citep{johnston_2019}. In the same vein, \cite{blueshear_bigwood} assumed null alignment for blue galaxies with an alignment uncertainty for each tomographic bin informed by redshift-analogous, star-forming, direct measurements.
Our COSMOS demonstration highlights an additional consideration, if treated conservatively: even when direct measurements constrain the IA of a particular galaxy population, the corresponding prior for a photometric sample can remain substantially broader if a significant fraction of its galaxies lies outside the calibrated demographic space. For our \textsc{blue} sample, this fraction is $32\%$, yielding a final prior of $A_{\mathrm{tot,blue}}=0.01\pm1.10$ under our conservative treatment of uncalibrated galaxies -- a broader prior than has generally been adopted in previous literature for star-forming samples.

The same framework can inform source selection. Rather than optimizing solely for galaxy number density, future analyses could balance statistical precision against IA calibration coverage, retaining populations for which precise priors are available.
Our tailored tomographic example illustrates the potential of this approach, although without an optimization for cosmological precision. More generally, the deep-field infrastructure used for photometric-redshift calibration provides a natural route toward jointly characterizing source redshift distributions and IA. Such an approach could help address the known interplay between redshift and IA nuisance parameters in cosmological inference \citep[e.g.,][]{fischbacher_redshift_ia_interplay,ia_photz_degen, johnston_6x2pt}.

\textbf{(iii) Expanding the IA measurement library}: The framework presented here is immediately applicable to current and upcoming weak-lensing surveys, including DES Y6, Rubin, Euclid, and Roman, which all probe different galaxy populations at different cosmic epochs \citep[e.g.,][]{openuniverse,roman_sam_galaxies,euclid_flagship}. Its predictive power will naturally improve as the library of direct IA measurements expands and their precision increases through surveys on multiplexed instruments like DESI/DESI-II \citep{desi_instrument,desi_ii_specs5}, PFS \citep{pfs}, and 4MOST \citep{4most}, and as the dependence of IA on additional galaxy properties, including morphology, environment, and halo mass, becomes better understood. As discussed, future work will sidestep the modeling correction described in App.~\ref{sec:alternate} by performing these direct measurements themselves in an astrophysically weighted capacity optimized for a given survey's selection and adapted to their respective shape measurement methods.

Equally importantly, the framework highlights where the current observational landscape remains incomplete. Regions of parameter space such as low-mass galaxies, high-redshift star-forming populations, and galaxy environments with under-sampled number densities remain undefined. Rather than simply motivating additional IA measurements, this framework identifies which galaxy populations currently limit cosmological analyses, providing a roadmap for future observational efforts. As more of this space is filled in, extrapolation between direct measurement populations that bookend alignment behavior will become more well motivated. With IA being a barrier to the use of small-scale correlation measurements, sample selection informed by astrophysical properties or the priors developed in this work is one avenue among many analysis methodologies to mitigate alignment \citep[e.g.,][]{heymans_2004_ia_mitigation,ia_mitigation_null_boost,ia_3x2,self_calibration_i,kidsxcmb_ia,bera_ia_self_calib_nl,lensing_counter_terms_smallscales,desi_steel,johnston_6x2pt}. Some of these methods (e.g., self-calibration) have the potential to provide \textit{indirect} measurements of alignment that we could also leverage into our empirical framework, addressing hard-to-measure populations without exposure-expensive spectroscopic campaigns.

Ultimately, realizing the full cosmological potential of next-generation weak-lensing surveys -- that will be characterized by fainter, higher redshift populations than ever before -- will require not only that increasingly sophisticated physical models of intrinsic alignments are developed, but also that those models are informed by priors appropriate to the unprecedented galaxy demographics to which they are applied.

\renewcommand{\arraystretch}{1.5} 
\begin{table}
    \centering
    \resizebox{\linewidth}{!}{
\begin{tabular}{c c c c c c}
	\hline
	Sample & $n_{\mathrm{gal}}$ & $f_u$ & $\overline{A}_{\mathrm{uncal}}$ & $\overline{A}_s$ & Final Prior, $\overline{A}_{\mathrm{tot}}$\\
	\hline
	\textsc{full} & 1.146 $\times 10^5$ & 0.33 & $\mathcal{U}(-5, 5)$ & $0.27_{-0.26}^{\ \ 0.26}$ & $0.17_{-1.14}^{\ \ 1.14}$ \\
    &   &    & $\mathcal{U}(-3,3)$ &   & $0.18_{-0.67}^{\ \ 0.68}$ \\
	\hline
    \textsc{blue} & 8.614 $\times 10^4$ & 0.32 & $\mathcal{U}(-5, 5)$ & $0.00_{-0.30}^{\ \ 0.30}$ & $0.01_{-1.09}^{\ \ 1.11}$ \\
     &   &   & $\mathcal{U}(-3,3)$ &   & $0.00_{-0.66}^{\ \ 0.66}$ \\
	\textsc{red} & 2.467 $\times 10^4$ & 0.57 & $\mathcal{U}(-5, 5)$ & $1.12_{-0.25}^{\ \ 0.25}$ & $0.47_{-1.94}^{\ \ 1.94}$ \\
     &   &   & $\mathcal{U}(-3,3)$ &   & $0.48_{-1.16}^{\ \ 1.16}$ \\
	\textsc{tomo} & 1.737 $\times 10^3$ & 0.02 & $\mathcal{U}(-5, 5)$ & $1.52_{-0.25}^{\ \ 0.25}$ & $1.48_{-0.25}^{\ \ 0.25}$ \\
     &   &   & $\mathcal{U}(-3,3)$ &   & $1.48_{-0.24}^{\ \ 0.25}$ \\
	\hline
\end{tabular}}

    \caption{The tailored priors, $\overline{A}_\mathrm{tot}$, for a variety of photometric selections in the COSMOS field. The uncalibrated fraction of each sample, $f_u$, and the informed prior on the calibrated component of each sample, $\overline{A}_s$ are reported with their 68\% intervals. Final priors are reported as the mean amplitude (of the distribution of means) alongside their 68\% confidence intervals -- though it should be noted that for many samples these are highly non-Gaussian, as explored in App.~\ref{app:uncal_dists}. See Fig.~\ref{fig:per_gal_prior} for more detail on each sample and for the \textsc{full} sample see App.~\ref{app:full}. The above results assume for fixed alignment across each direct measurement sample; for a simple analytic correction that accounts for gradients within the populations, see App.~\ref{sec:alternate}. }
    \label{tab:prior_outputs}
\end{table}

\section{Conclusions}\label{sec:concl}
Intrinsic alignments present a major challenge for precision weak-lensing cosmology \citep[e.g., ][]{amon_2021,secco_2021,decade_v,unions,Wright_2025, desy6_cosmic_shear, chenderose2024}; restrictive models risk biasing cosmological inference, while more flexible models can substantially degrade constraining power, introduce projection effects, and drive sensitivity to poorly-constrained nuisance parameters and their prior choices. We have developed a practical framework for constructing IA priors tailored to the astrophysical properties of weak-lensing source samples, using existing direct measurements of galaxy alignments. Our approach combines direct IA measurements with the joint distributions of inferred astrophysical properties (e.g., stellar mass and specific star-formation rate) for both spectroscopic calibrator samples and photometric source populations. By explicitly identifying the fraction of source galaxies not covered by existing measurements, we distinguish between empirically calibrated IA information and the assumptions adopted for uncalibrated populations.
In this work, we have:
\begin{itemize}
    \item \textbf{Characterized a demographic library of direct IA measurements} on alignment, compiling them into a common framework (Sec.~\ref{sec:library}) across a range of galaxy populations and rendering their joint distributions in stellar mass (a halo-mass proxy) and specific star-formation rate  (a color proxy). This provides an empirical basis for relating measured alignments to photometric source samples with different galaxy demographics, and allows us to better understand the impact of population gaps --  dominated by low-mass galaxies -- in existing measurements.
    \item \textbf{Developed a model-agnostic prescription for constructing tailored-IA priors} (Sec.~\ref{sec:methods}) that relies only on the direct measurements and the demographic overlap between those populations and a target photometric population to be calibrated. Crucially, the method explicitly identifies galaxies outside the effective coverage of existing measurements and assigns them a broad, uninformative prior rather than extrapolating measured alignment amplitudes.
    \item \textbf{Demonstrated the framework using case studies} of galaxies in the Dark Energy Survey COSMOS field (Sec.~\ref{sec:results}), deriving IA priors for \textsc{blue}, \textsc{red}, and \textsc{full}  selections. We find that approximately one-third of the blue sample and more than half of the red sample lack effective direct IA calibration -- territory where other approaches to alignment are forced to extrapolate, but where our method adopts a deliberate, uninformative prior. Consequently, even relatively precise measurements of the calibrated populations can yield broad priors for the complete source samples.
    \item \textbf{Showcased the potential for  integration of the tailored-IA method with common infrastructure for photometric redshift inference} (Sec.~\ref{sec:som})  to map the available IA information across photometric color space. We identify which direct measurements calibrate different source populations and where coverage remains incomplete. Moreover, as a roadmap for future analyses, we perform a simple, tailored tomographic selection that has $98\%$ calibration coverage (though low statistical power), yielding an IA prior approximately $4.5\times$ narrower than that of the full sample. This lays groundwork for analyses that balance sample selection against cosmological precision. 
\end{itemize}

With next-generation weak-lensing observations from Rubin, Euclid, and Roman already underway \citep{lsst_dp1, lsst_dp2, euclid_q1}, increasing statistical precision demands correspondingly robust treatments of intrinsic alignments. At the same time, massively multiplexed spectroscopic surveys will enable comprehensive direct IA measurements across an increasingly diverse range of galaxy populations \citep[e.g.,][]{4most, pfs, desi_ii_specs5}, while advances in forward modeling are improving the astrophysical characterization of photometric source samples \citep[e.g.,][]{galsbi-sps,galsbi,pop-cosmos,popsed}. The framework developed here provides a way to connect these advances, translating the growing empirical knowledge of galaxy alignments into priors tailored to the galaxies used in cosmological analyses. Our results highlight that robust IA mitigation requires not only increasingly sophisticated models, but also empirical constraints appropriate to the source populations being analyzed. Both the expansion of direct and indirect IA measurements, especially for populations that remain poorly observed, and accurate constraints on astrophysical galaxy properties for survey-specific populations will be essential to realizing the full cosmological potential of next-generation weak-lensing data.



\section*{Data Availability}
No new data were observed or generated for this paper. Some DESI Y1 direct observations were measured in different galaxy splits than reported in \cite{siegel_2025}, in order to optimize their constraining power. These updated $w_{g+}$ measurements are made available alongside additional data products for this work at \url{https://jamiemccullough.github.io/data/tailoredIA/}. 

\section*{Acknowledgments}
We thank Anand Raichoor for providing additional data products detailing astrophysical properties for the eBOSS spectroscopic sample, and we thank Gary Bernstein for engaging with very useful discussions on the paper that resulted in improvements to the method.

This work benefited from the resources and support provided by the echo-IA collaboration (\url{https://echo-ia.org}), and the additional discussions that were held with alignment experts at the Utrecht UNBIAS\footnote{\url{https://echo-ia.github.io/UtrechtIA/}} echo-IA workshop, partially funded by the European Research Council (ERC) and the Dutch Research Council (NWO). 

This publication is funded in part by the Gordon and Betty Moore Foundation through Grant GBMF13897.

BJ acknowledges support by the ERC-selected UKRI Frontier Research Grant EP/Y03015X/1. DNG acknowledges support from the European Research Council (ERC) under the European Union’s Horizon 2020 research and innovation program with Grant agreement No. 101053992


\bibliographystyle{mnras}
\bibliography{references}




\appendix

\section{A catalog of direct measurements of intrinsic alignment}\label{app:overlap_dm}

While the direct measurements used in this work have alignments and astrophysically properties summarized in Fig.~\ref{fig:diagnost}, we elaborate on the sources of those values and catalogs in Table~\ref{tab:direct_meas}, enumerating each sample, their inference for NLA amplitude $A_{\mathrm{IA}}$, the publication detailing their source of precise redshifts, shape measurement, and IA direct measurement (and relevant target selection and large-scale structure catalogs if applicable). We further detail any selections made on specific-star-formation rate or stellar mass. As mentioned in Sec.~\ref{sec:library}, this paper is limited to public direct measurements that have catalogs with these properties readily available. Future work will expand on this.

\begin{figure}
    \centering
    \includegraphics[width=\linewidth]{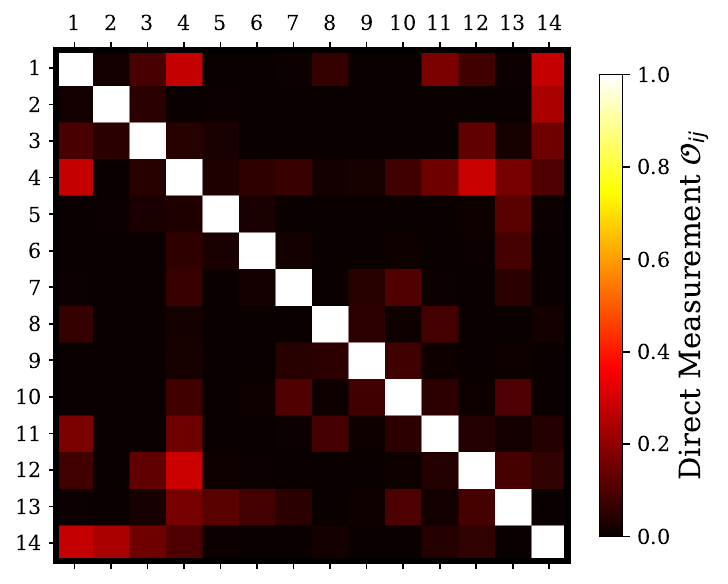}
    \caption{The 2D overlap coefficient $\mathcal{O}_{ij}$ (in the sSFR and $M_*$ plane) between each of the galaxy populations with direct measurements on alignment (numbered in Table~\ref{tab:direct_meas}). As there is remarkably small overlap between the different measurement samples, this work  only accounts for off-diagonal overlap for samples that are non-detections. As this is a measure of the similarity of the astrophysical populations, it is similar to a positive correlation coefficient and can be used to estimate a covariance matrix.}
    \label{fig:measurement_overlap}
\end{figure}

While Fig.~\ref{fig:diagnost} shows that these samples, for the most part, populate unique portions of the sSFR-$M_*$ plane, there is some overlap between the astrophysical properties between samples. We compute the overlap coefficient from Eqn.~\ref{eqn:overlap} between the samples (as numbered in Table~\ref{tab:direct_meas}) and depict them in Fig.~\ref{fig:measurement_overlap}. It can be noted that these values are quite low, though there is some overlap ($\sim0.3$) between the bluer samples, e.g., DESI BGS (12) + PAUS Blue (4), and eBOSS ELGs (1) + DESI ELGs (14). 

We treat the overlap matrix $\mathcal{O}_{ij}$ as an expected correlation coefficient between the two samples that allows for estimation of the covariance of their alignment estimates given their respective $A_{\mathrm{IA}}$ measurement uncertainties, $\sigma$. If $\mathcal{O}_{ij}$ is positive and semi-definite then the covariance follows as
\begin{equation}
    C_{ij} = \mathcal{O}_{ij}\ \sigma_i\ \sigma_j.
\end{equation}

This covariance computation step is complicated in this work by the lack of matching-astrophysical-property weighting between direct measurements (discussed in Sec.~\ref{sec:practical}), as nonlinear gradients in $A_{\mathrm{IA}}$ across the samples likely exist and a high overlap coefficient may not accurately capture the true correlation of the measurements. Therefore, we employ it only when $i,j$ are both non-detections but set off-diagonal covariance elements for non-zero aligning samples to zero in this work. This happens to work well for our samples, as the most significant overlap between measurements is not for aligning populations.

\renewcommand{\thefootnote}{\fnsymbol{footnote}}
\setlength{\tabcolsep}{2pt}

\begin{table*}
    \centering
    \resizebox{\textwidth}{!}{\begin{tabular}{c c c c c c c c}
    \hline
    \hline
     & Sample &$A_{\mathrm{IA}}$& Redshift Source & Shape Source & sSFR \& $M_*$ & IA Reference & \makecell{Target Selection\\LSS Reference}\\
    \hline
    \hline
    1 & eBOSS ELG & $0.07\ ^{+0.32}_{-0.42}$ & eBOSS Emission Line Galaxies & DES Y3 & \textsc{fast}\footnotemark[1], \cite{fast_method_kriek_2009}\footnotemark[2] & \cite{samuroff_2023}& \makecell{\cite{eboss_elg_final,eboss_elg_lss_bao}\\\cite{eboss_elg_growthrate}}\\
    \hline
    2 & PAUS Blue bin 1  & $0.98\ ^{+1.98}_{-1.98}$& \makecell{PAUS photo-$z$s\\\cite{paus_photoz}} & CFHTLenS, KiDS & \makecell{\textsc{cigale}, \cite{paus_stellar}\\$\rm{log_{10}}(M_*) < 9.61$} & \cite{paus_direct}& \\
    3 & PAUS Blue bin 2  & $1.61\ ^{+0.91}_{-0..91}$& " " & " " & $9.61 < \rm{log_{10}}(M_*) < 10.22$& " "&  \\
    4 & PAUS Blue bin 3  & $-0.22\ ^{+1.38}_{-1.44}$& " " & " " & $\rm{log_{10}}(M_*) > 10.22$ & " "&  \\
    \hline
    5 & PAUS Red bin 1  & $0.95\ ^{+1.19}_{-1.21}$& \makecell{PAUS photo-$z$s\\\cite{paus_photoz}} & CFHTLenS, KiDS & \makecell{\textsc{cigale}, \cite{paus_stellar}\\$\rm{log_{10}}(M_*) < 10.48$} & \cite{paus_direct} & \cite{}\\
    6 & PAUS Red bin 2  & $4.30\ ^{+1.61}_{-1.55}$& " " & " " & $10.48 <\rm{log_{10}}(M_*) < 10.81$ & " " & \cite{}\\
    7 & PAUS Red bin 3  & $6.07\ ^{+1.61}_{-1.54}$& " " & " " & $\rm{log_{10}}(M_*) > 10.81$ & " " & \\
    \hline
    8 & DESI LRG bin 1  & $3.19\ ^{+1.15}_{-1.18}$& \makecell{DESI Y1 Luminous Red Galaxies\\\cite{desi_y1}} & DES Y3 + KiDS-1000 & \makecell{\textsc{cigale}, \cite{siudek_cigale_desiy1}\\$\rm{log_{10}}(M_*) > 11.20\ \&\ \rm{log_{10}}(sSFR) > -11.20$} & \makecell{\cite{siegel_2025}\\Jeffrey et al. (in prep)} & \makecell{\cite{desi_lrg_targets}\\\cite{Ross_2025}}\\
    9 & DESI LRG bin 2  & $4.70\ ^{+0.84}_{-0.83}$& " " & " " & $\rm{log_{10}}(M_*) > 11.20\ \&\ \rm{log_{10}}(sSFR) < -11.20$ & " " & " "\\
    10 & DESI LRG bin 3  & $1.74\ ^{+0.49}_{-0.44}$& " " & " " & $\rm{log_{10}}(M_*) < 11.20\ \&\ \rm{log_{10}}(sSFR) < -11.20$ & " " & " "\\
    11 & DESI LRG bin 4  & $-1.02\ ^{+0.91}_{-0.89}$& " " & " " & $\rm{log_{10}}(M_*) < 11.20\ \&\ \rm{log_{10}}(sSFR) > -11.20$ & " " & " "\\
    \hline
    12 & DESI BGS bin 1  & $-0.99\ ^{+0.57}_{-0.60}$& \makecell{DESI Y1 Bright Galaxy Survey\\\cite{desi_y1}} & SDSS + DES Y3 + KiDS-1000 & \makecell{\textsc{cigale}, \cite{siudek_cigale_desiy1}\\$\rm{log_{10}}(M_*) > 10\ \rm{and}\ \rm{log_{10}}(sSFR) > -11.20$} & \makecell{\cite{siegel_2025}\\Jeffrey et al. (in prep)} & \makecell{\cite{desi_bgs_targets}\\\cite{Ross_2025}}\\
    13 & DESI BGS bin 2  & $3.27\ ^{+0.63}_{-0.66}$& " " & " " & $\rm{log_{10}}(M_*) > 10\ \rm{and}\ \rm{log_{10}}(sSFR) < -11.20$ & " " & " "\\
    \hline
    14 & DESI ELG  & $-0.85\ ^{+0.66}_{0.66}\ $\footnotemark[3] & \makecell{DESI Y1 Emission Line Galaxies\\\cite{desi_y1}} & HSC Y3 & \textsc{cigale}, \cite{siudek_cigale_desiy1} & \makecell{\cite{siegel_2025}\\Jeffrey et al. (in prep)} & \makecell{\cite{desi_elg_targets}\\\cite{Ross_2025}}\\
    \hline
    \end{tabular}}
    \caption{A summary of the consolidated direct measurements of intrinsic alignments used in this work, and the relevant references from which they derive their position (redshift), photometry (shape), and target selection. Additionally noted, when relevant, are the data products and catalogs used for large-scale structure (LSS) measurements. While many other measurements have been made, the above observations have available estimates of their specific star formation rates (sSFR) and stellar masses ($M_*$) -- essential for placing them in our inference plane. Note that multiple surveys have performed their SED fitting with \textsc{cigale} \citep{cigale_boquien}.}
     \label{tab:direct_meas}
\end{table*}

\footnotetext[1]{The sSFR measurements from the \textsc{fast} algorithm were coarsely discretized and not reported in the original publication. Gaussian smoothing was applied in our proof-of-concept analyses to its distribution to make it more tractable, but future work would perform independent SED fits to these galaxies.}

\footnotetext[2]{\url{https://data.sdss.org/datamodel/files/EBOSS_LSS/catalogs/DR16/eBOSS_ELG_full_ALLdata.html}}

\footnotetext[3]{The DESI ELGs were fit to smaller scales than other measurements ($>2$ Mpc/h) to improve the constraining power on the null detection.}

\renewcommand{\thefootnote}{\arabic{footnote}}

\section{Uncalibrated fraction and the shape of the tailored prior}\label{app:uncal_dists}

To build intuition for the contribution of the uncalibrated sample to the prior set on the overall mean of the alignment amplitude, this section demonstrates a simple example with many choices of the uncalibrated fraction, $f_u$. For a calibrated sample whose means are drawn from $\overline{A}_s\sim \mathcal{N}(\mu,\sigma)$ with example parameters $\mu=1.5$ and $\sigma=0.25$, and an uncalibrated sample whose means are drawn from $\overline{A}_{\mathrm{uncal}}\sim \mathcal{U}(-5,5)$, we plot their combination for 1 million draws following Eqn.~\ref{eqn:combine_frac}. 

\begin{figure}
    \centering
    \includegraphics[width=\linewidth]{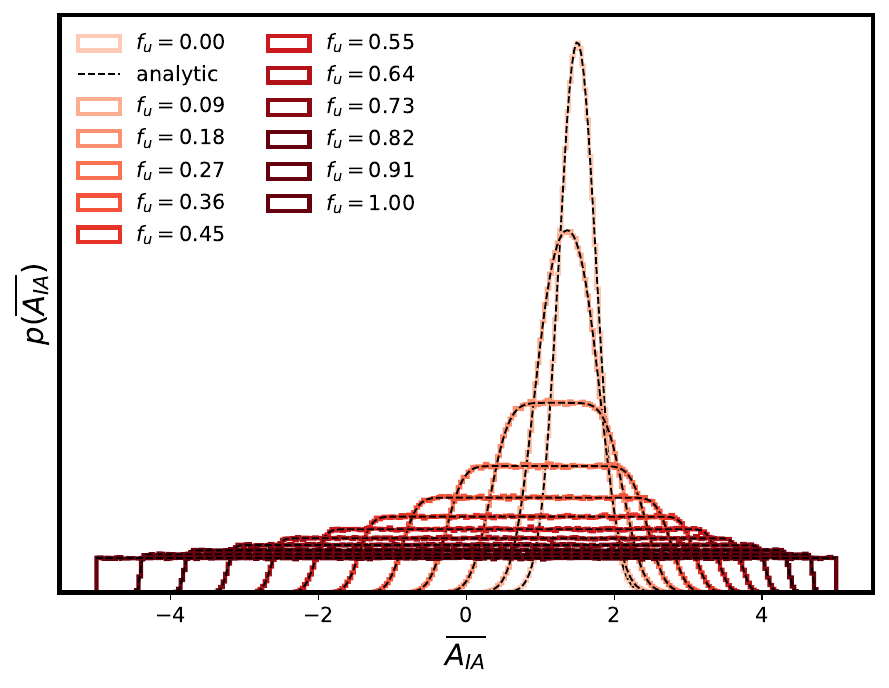}
    \caption{An example of the shapes the distribution might take, when it describes the mean of alignment from a calibrated population (with mean modeled by the Gaussian, $f_u = 0$) and an uncalibrated population (with mean drawn from a flat, uniform distribution, $f_u = 1$). The uncalibrated fraction $f_u$ determines the exact mixture. In black dashed over each line is the analytic, closed form described by Eqn.~\ref{eqn:analytic}, where the colored lines represent the distribution sampled with one million draws. }
    \label{fig:toymodel}
\end{figure}

The histogrammed results are in Fig.~\ref{fig:toymodel} (color) for $0 < f_u < 1$, showcasing the dominant regimes of either component and the intermediate form of a flat topped distribution with short tails. For broader choices of $\sigma$, the tails become more pronounced. From the other case studies in this work (see Fig.~\ref{fig:per_gal_prior}) and the prescription of marginalizing over Gaussian measurement uncertainties for each direct measurement amplitude of alignment, the assumption of normally distributed $\overline{A}_s$ is reasonable for this work. With that assumption, our analytic solution from Eqn.~\ref{eqn:analytic} is a perfect description of these draws (see black dashed lines).

\section{Method validation, analytic priors, and gradient corrections}\label{sec:alternate}

Rather than the empirical approach explored in this work, other data-driven priors for alignment have favored analytic forms for alignment that are primarily a function of luminosity \citep{unions, hervas_peters_2024_unions} or halo-mass \citep{Wright_2025,fortuna_2025}. These empirical formulations extrapolate to lower-mass galaxies than have been directly measured. Typically, power-laws are chosen to describe the alignment behavior (or broken power-laws), e.g., for halo mass $M_h$,

\begin{equation}
    A_{\rm{IA}}(M_h) = A\left(\frac{M_h}{M_0}\right)^\gamma,
\end{equation}
where $A=5.74\pm0.32$, $M_0 = 10^{13.5}h^{-1}M_\odot$ and $\gamma=0.44\pm0.04$ in \cite{fortuna_2025}. 

\subsection{Analytic Modeling of our Observations}
As our work uses stellar mass as our overall mass and luminosity proxy and further includes a trend for color (captured in specific star-formation rate), we can fit an analytic model of alignment of the following form,

\begin{equation}\label{eqn:analytic_model}
    A_{\mathrm{IA}}(M_*,\mathrm{sSFR}) = A \left(\frac{M_*}{M_0}\right)^{\alpha} f_c(\mathrm{sSFR}),
\end{equation}
where the exponent for stellar mass and the color term, $f_c$, are described piecewise, 
\begin{equation}
    \alpha = \begin{cases} 
\alpha_1 & \text{if } M_* < M_0, \\ 
\alpha_2 & \text{if } M_* \ge M_0 
\end{cases}\ \ \mathrm{and} \\
\end{equation}
\begin{equation}\label{eqn:colorterm}
f_c(\mathrm{sSFR}) = \begin{cases} 0 &\text{if sSFR} > R_0, \\ \mathrm{log}_{10}\mathrm{sSFR} - \mathrm{log}_{10}R_0 & \text{if sSFR} < R_0.
\end{cases}
\end{equation}
Eqn.~\ref{eqn:colorterm} is chosen to set star-forming galaxies above some threshold, $R_0$, to zero alignment and then contribute linearly for redder, quenched galaxies. While simple, more sophisticated treatment of the color term did not significantly improve our $\chi^2$.

We fit this model with wide, uniform priors to the direct measurements described in Table \ref{tab:direct_meas}, using their full population distributions in stellar mass and sSFR, $p_i(M_*,\mathrm{sSFR})$, where the analytic model is computed over the same grid that $p_i$ is before integrating. While fitting, we set negative measured amplitudes to zero and retain their reported errors. Our priors and posteriors of the fit are described in Table \ref{tab:prior_analytic}. As mentioned in Sec.~\ref{sec:practical}, we anticipate systematic offsets in the astrophysical property inference for the different surveys. For this reason, we also could incorporate small shifts as free parameters to the $p_i$ distributions, coherent across their respective surveys: $(\Delta \mathrm{log}_{10} M_{*})_\mathrm{DESI}, (\Delta \mathrm{log}_{10} M_{*})_\mathrm{PAUS}, (\Delta \mathrm{log}_{10} M_{*})_\mathrm{eBOSS}$. In practice, eBOSS contributes one sample that is a non-detection, resulting in an unconstrained shift parameter, while only one shift between DESI or PAUS would be necessary to bring their systems into agreement. We retain the DESI shift and discard the others. While sSFR is also uncertain, incorporating additional shifts does not improve our $\chi^2$. It is likely that these shifts are non-linear, rendering this model a non-rigorous example with simplified systematic treatment. Since we only incorporate the correction to stellar mass,
\begin{align}
    \mathrm{log}_{10}M_*' &= \mathrm{log}_{10}M_* + (\Delta \mathrm{log}_{10} M_{*}).
\end{align}

\begin{table}
    \centering
    \begin{tabular}{c c c c}
        \hline
        Parameter & Prior & Posterior & Max-Like. \\
        \hline
        $A$ & $\mathcal{U}(-10,10)$ & $1.33^{+1.34}_{-0.63}$ & 0.81 \\
        log$_{10}M_0$ & $\mathcal{U}(9.5,11.5)$ & $10.97^{+0.35}_{-0.55}$ & 10.94 \\
        log$_{10}R_0$ & $\mathcal{U}(-12,-8)$ & $-9.78^{+0.48}_{-0.62}$ & -9.49\\
        $\alpha_1$ & $\mathcal{U}(-3,3)$ & $0.34^{+0.52}_{-0.39}$ & 0.00\\
        $\alpha_2$ & $\mathcal{U}(-3,3)$ & $0.71^{+0.79}_{-0.97}$ & 1.87\\
         \hline
        $(\Delta \mathrm{log}_{10} M_{*})_\mathrm{DESI}$ & $\mathcal{U}(-0.6,0.6)$ & $-0.28^{+0.27}_{-0.17}$ & -0.26 \\
         \hline
    \end{tabular}
    \caption{The prior and posterior values for the model described in Eqn.~\ref{eqn:analytic_model}. When fitting the observations made in Table~\ref{tab:direct_meas}, the model has a $\chi^2$ of 16.3, and a reduced $\chi^2_{\mathrm{red}}$ of $\sim1.8$. The maximum-likelihood value from the chain is reported naively, without a dedicated optimizer.}
    \label{tab:prior_analytic}
\end{table}

Running MCMC chains to minimize the $\chi^2$ between model predictions and direct observations using the \texttt{emcee} library \citep{emcee}, we find a best-fit of $\chi^2_{\mathrm{red}}\equiv\chi^2/\mathrm{dof} = 16.3/9\approx1.8$, similar to the goodness-of-fit values described in \cite{siegel_2025} (their Sec.~5), though we fit to more direct measurements and include alignment measures from blue galaxies. 

\subsection{Method Validation}
This model could be a basis for an analytic prior -- if we used it for our \textsc{blue} sample we would infer $A_{\mathrm{IA}} = 0.01^{\ +0.05}_{-0.00}  \ (0.072)$ (model best-fit), $A_{\mathrm{IA}} = 0.76^{\ +0.50}_{-0.20}\ (1.45)$ for \textsc{red}, and $A_{\mathrm{IA}} = 0.14^{\ +0.15}_{-0.05}\ (0.32)$ for \textsc{full}. These all fall in agreement with the 68\% intervals of our empirical results in Table~\ref{tab:prior_outputs}. We do see minor disagreement $(\sim2\sigma)$ for the \textsc{tomo} selection, where both the measurements and the model have the most extreme alignment behavior, with the model predicting $A_{\mathrm{IA}} = 2.27^{\ +0.52}_{-0.44}\ (2.81)$.

Furthermore, given this analytic model, we can validate our pipeline by mocking up direct observations. If we take the posteriors in Table~\ref{tab:prior_analytic}, and integrate it over the same $p_i(x,y)$ distributions we have for our real data, we produce synthetic alignment amplitudes for each direct measurement. The result is depicted as the $A$ \textsc{model} column in Table~\ref{tab:aia_corrected}. If we assume the mean posterior model as a perfect description of alignment in some universe, our mean $A$ \textsc{model} values are exact. If we attribute to them some arbitrary, small uncertainty of $\pm0.1$ and repeat our tailored-IA inference, our posterior on alignment should encompass the analytic prior for each sample in the previous paragraph. Our synthetic measurements yield $A_{\mathrm{IA}} = 0.07^{\ +1.11}_{-1.08}$ for \textsc{blue}, $A_{\mathrm{IA}} = 0.52^{\ +1.91}_{-1.94}$ for \textsc{red}, and $A_{\mathrm{IA}} = 0.23^{\ +1.14}_{-1.13}$ for \textsc{full}. These agree well with the model expectation for all major samples in our work with the exception of the strongly aligning \textsc{tomo} sample, where the modeled direct measurements, when used without correction, yield $A_{\mathrm{IA}} = 1.50^{\ +0.07}_{-0.07}$. This disagreement can be attributed to the strong gradient behavior of alignment across these measurements, as our photometric sample to be calibrated has galaxies with similar properties to the measurements but very different relative abundances.

As we mention this obstacle in the main text, we can validate that our recommended weighting procedure for $w_{g+}$ measurements in future work will result in an unbiased alignment estimate with our analytic model under ideal conditions. For each photometric sample, we produce an `ideal' set of calibration measurements that together span the full parameter space and, for their respective region, match the astrophysical distribution of the target sample. For this test, we cut the sSFR-$M_*$ plane into 10 x 10 blocks, and consider each block as sourcing an independent calibrator measurement and compute its alignment with Eqn.~\ref{eqn:ia_mean} and ~\ref{eqn:analytic_model}. 

\begin{figure}
    \centering
    \includegraphics[width=\linewidth]{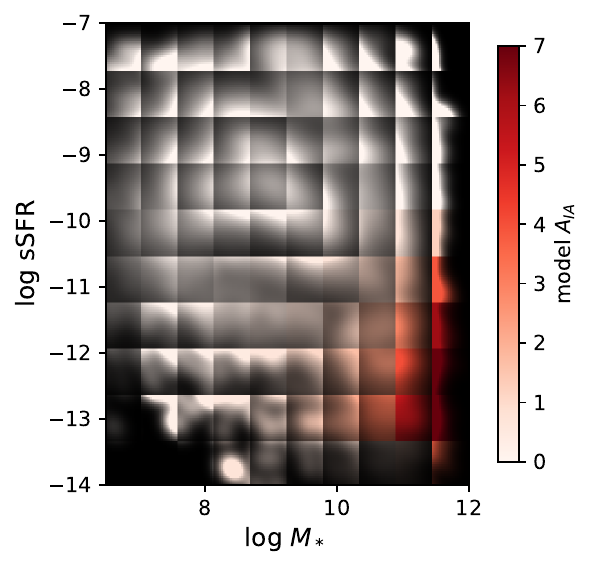}
    \caption{The $p_i(x,y)$ distributions of an ideal synthetic calibrator sample (100 subpopulations) for our \textsc{full} COSMOS selection, informed by our analytic model and weighted in each domain to match its corresponding photometric distribution. Black regions depict areas where $p_i(x,y) = 0$, and color indicates the mean alignment for these samples from our model. Similar ideal samples were generated for the \textsc{blue}, \textsc{red}, and \textsc{tomo} selections in this work.}
    \label{fig:synthetic}
\end{figure}

This gives us 100 independent synthetic direct measurements from our analytic model for each of our photometric samples. An example of these measurements and their $p_i(x,y)$ distributions to match the \textsc{full} sample can be seen in Fig.~\ref{fig:synthetic}, colored by their mean alignment amplitude. With an assumed per-measurement uncertainty of $\pm0.05$ on the amplitudes (not encompassing any model uncertainty), running through our tailored-IA pipeline we recover $A_{\mathrm{IA}} = 0.01^{\ +0.02}_{-0.02}$ for \textsc{blue}, $A_{\mathrm{IA}} = 0.74^{\ +0.17}_{-0.17} $ for \textsc{red}, $A_{\mathrm{IA}} = 0.14^{\ +0.01}_{-0.01}$ for \textsc{full}, and $A_{\mathrm{IA}} = 2.27^{\ +0.27}_{-0.27}$ for \textsc{tomo} -- all in excellent agreement with our analytic expectation. Some additional uncertainty enters this estimate because galaxies that fall outside the chosen astrophysical limits are uncalibrated. 
This test demonstrates that the tailored method works well (a) if all trends in alignment are captured in the chosen astrophysical axes and (b) if the galaxy abundances in the direct measurements match the target sample on these axes. The latter condition is not true for the results in the main text of our work as the direct measurements are sourced from many surveys and analysis teams -- therefore requiring a first-order correction and an assumed alignment parametrization.

\subsection{An Estimate for an Alignment-Gradient Correction}
In light of the unavoidable mismatch of existing measurements to our photometric samples, we can use our analytic form to estimate an approximate correction to the measured alignment amplitudes that accounts for gradients within individual direct measurements -- which future inferences should avoid by performing direct measurements that astrophysically weight their samples like the photometric sample (as demonstrated above). Unless the amplitude is constant across the sample, our overlap metric will potentially attribute significant calibration power from galaxies that are not well represented by a given mean measurement. The following is not a recommended prescription for prior treatment, and only meant to illustrate the magnitude of potential offsets between a raw direct measurement and one optimized for a given photometric sample.  

Our procedure for performing a first-order correction takes the best-fit parameters for the model described in Table~\ref{tab:prior_analytic}, where we
\begin{itemize}
    \item Compute a linear offset, $L$, for each direct measurement that minimizes any residual between the model prediction and the direct measurement amplitude. This reduces the impact of the outlier measurements and allows us to focus on the gradient of alignment across the astrophysical plane that is most relevant to a given direct measurement.
    \item For each photometric sample to be calibrated, we compute $\mathrm{min}(p_i,p_{\mathrm{phot}})$ in our sSFR-$M_*$ plane (as seen in Eqn.~\ref{eqn:overlap}).
    \item We produce an estimate for the modeled amplitude of alignment of the direct measurement population when restricted to the overlap with the photometric sample by integrating that distribution with the fit model:
    \begin{equation}
        A_i' = \frac{\iint \mathrm{min}\left[p_i(x,y),p_{\mathrm{phot}}(x,y)\right]\left(A_{\mathrm{IA}}(x,y)+L_i\right)dxdy}{\iint \mathrm{min}\left[p_i(x,y),p_{\mathrm{phot}}(x,y)\right]dxdy}
    \end{equation}
    \item We repeat our inference with these $A_i'$ values, tabulated for each photometric sample in Table~\ref{tab:aia_corrected} and inferred for the photometric galaxies in each sample that overlap direct measurements, $\overline{A}_s'$, and for the total sample, $\overline{A}_{\mathrm{tot}}'$, with the mean alignment of uncalibrated galaxies drawn from $\mathcal{U}(-5,5)$ as in the fiducial analysis.
\end{itemize}

\begin{table}
    \centering
    \begin{tabular}{c c c c c  c}
	\hline
	IA Sample & $A${
    \textsc{model}} & $A'$\textsc{full} & $A'$\textsc{blue} & $A'$\textsc{red} & $A'$\textsc{tomo} \\
	 \hline
	eBOSS ELG& $0.09^{ + 0.12}_{- 0.07}$&0.13* & -0.01* & 0.79 & 0.40* \\
	PAUS Blue 1& $0.01^{+ 0.03}_{- 0.01}$& 0.95* & 0.93* & 1.08* & 1.10* \\
	PAUS Blue 2& $0.07^{+ 0.08}_{- 0.04}$& 1.72* & 1.56* & 2.04 & 2.04 \\
	PAUS Blue 3& $0.77^{+ 0.27}_{- 0.22}$& 0.26* & -0.76* & 1.02* & 0.83* \\
	PAUS Red 1& $0.91^{ + 0.36}_{- 0.35}$& 1.24* & 0.25* & 1.51* & 1.68* \\
	PAUS Red 2& $0.87^{ + 0.21}_{- 0.20}$& 7.99 & 7.13 & 8.06 & 8.22 \\
	PAUS Red 3& $1.19^{+ 0.31}_{- 0.29}$ & 9.74 & 8.32 & 10.03 & 9.49 \\
	DESI LRG 1& $1.37^{ + 2.09}_{ - 0.94}$ & 3.14 & 2.32* & 3.82 & 3.82 \\
	DESI LRG 2& $5.58^{+ 4.77}_{- 2.39}$& 4.80 & 3.65 & 5.10 & 5.23 \\
	DESI LRG 3& $3.55^{+ 1.25}_{- 0.96}$& 1.57 & 2.35 & 1.65 & 1.83 \\
	DESI LRG 4& $0.73^{ + 0.54}_{- 0.45}$& -0.21* & -0.53* & 0.32* & 0.25* \\
	DES BGS 1& $0.63^{+ 0.30}_{- 0.33}$& -0.25* & -0.51* & -0.07* & 0.08* \\
	DESI BGS 2& $1.73^{+ 0.38}_{ - 0.41}$& 3.68 & 2.67 & 3.61 & 3.61 \\
	DESI ELG& $0.05^{+ 0.06}_{ - 0.04}$& 0.02* & -0.03* & 0.33* & 0.20* \\
    \hline
    \textbf{$\overline{A}_s'$} & - &  $0.45^{\ +0.26}_{-0.26}$& $0.07^{\ +0.30}_{-0.30}$& $1.51^{\ +0.26}_{-0.25}$ & $1.84^{\ +0.25}_{-0.25}$\\
	\textbf{$\overline{A}_{\mathrm{tot}}'$} & - &$0.30^{\ +1.13}_{-1.12}$ & $0.06^{\ +1.09}_{-1.12}$ & $0.66^{\ +1.90}_{-1.88}$ &$1.81^{\ +0.25}_{-0.25}$ \\
	\hline
    \textbf{$\overline{A}_s'$*} & - & $0.28^{\ +0.26}_{-0.26}$ & $0.00^{\ +0.30}_{-0.30}$ & $1.51^{\ +0.26}_{-0.25}$ & $1.71^{\ +0.25}_{-0.25}$\\
	\textbf{$\overline{A}_{\mathrm{tot}}'$*} & - & $0.19^{\ +1.15}_{-1.13}$ & $0.01^{\ +1.08}_{-1.09}$& $0.72^{\ +1.91}_{-1.92}$& $1.68^{\ +0.25}_{-0.25}$\\
    \hline
\end{tabular}

    \caption{The corrected alignment amplitudes, $A'$, for each direct measurement population assuming the mean model parameters described in Table~\ref{tab:prior_analytic}, when limited to their respective overlap with a given photometric sample. These alignment amplitudes are rough estimates based upon the analytic model posteriors, intended to demonstrate the effect unaccounted-for gradients can have on the tailored inference. We repeat the inference pipeline for the main analysis exactly, inferring the calibrated sample prior mean, $\overline{A}_s'$*, and the total sample prior mean, $\overline{A}_{\mathrm{tot}}'$*, where we set any measurement consistent with non-detection to zero. Non-detections are marked with (*) given the reported direct measurement error. For completeness, we also compute the model-corrected alignment prior without this non-detection shift, as seen reported in $\overline{A}_s'$ and $\overline{A}_{\mathrm{tot}}'$.}
    \label{tab:aia_corrected}
\end{table}

\begin{figure}
    \centering \includegraphics[width=0.95\linewidth]{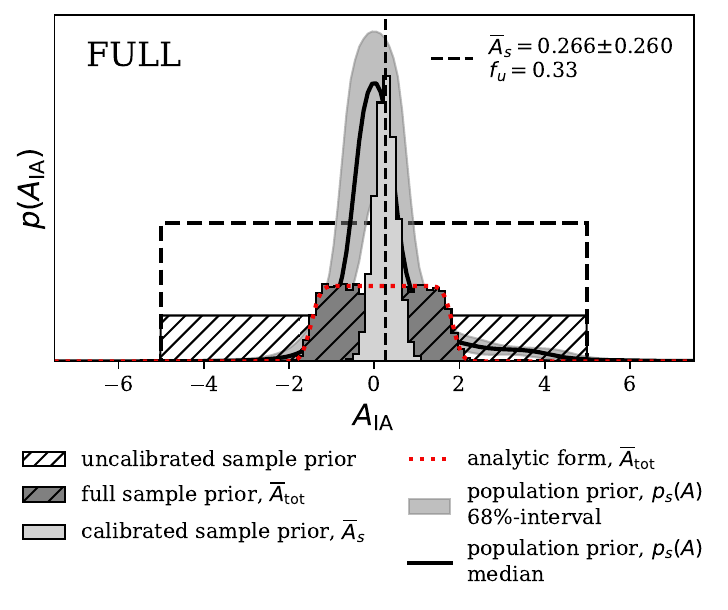}
    \caption{The full sample version of Fig.~\ref{fig:per_gal_prior}, where we plot the population-level distribution of alignment $p_s(A)$ (line) along with the 68\%-interval of their bootstrapped realizations (gray shaded). For these realizations we plot their means (light histogram) that represent the prior on alignment for the calibrated sample, $\overline{A}_s$. When combined with the flat prior set on the uncalibrated component of the population (hatched, visually weighted by $f_u$ within dashed region), we also display the final priors for the whole sample (dark histogram, hatched), $\overline{A}_{\mathrm{tot}}$. Note that relative normalizations are chosen for readability and are not physical.}
    \label{fig:full_result}
\end{figure}

We find, for all samples, final prior results whose 68\%-regions are consistent with our main analysis (summarized in Table~\ref{tab:prior_outputs}). For \textsc{blue} galaxies, this result is expected as our analytic model does not predict a gradient in alignment for star-forming galaxies, but rather a null signal. Another analytic model that predicted a trend for alignment in these galaxies could produce a meaningfully different result. This also sensibly holds for the \textsc{full} sample, which is dominated by those bluer galaxies. It is more impactful that we see consistent total results for the \textsc{red} and \textsc{tomo} samples, which contain galaxies where the modeled gradients would be most extreme, though this is in part due to the high uncalibrated fraction for the \textsc{red} sample. For the calibrated galaxies in \textsc{red}, we find a shift to higher expected alignment by $\sim 0.39$, which exceeds the 68\% interval of the non-gradient analysis in the main text. We also see similar upward shifts for gradient-aware treatment of the \textsc{tomo} sample, though the shift is less statistically significant.

Similarly in Table~\ref{tab:aia_corrected}, we explore the impact of assuming that insignificant detections (consistent with zero alignment within 2$\sigma$) have null alignment as we have done in the main text, and see minor shifts in the resulting priors when this assumption is in place (marked with *).

\section{Tailored priors for the full DES-COSMOS sample}\label{app:full}

While we report results for various subsets of this sample in Sec.~\ref{sec:results}, for completeness we also produce a prior with the proposed method for the uncut DES-COSMOS sample, \textsc{full}, here. This sample is dominated by the contributions from the \textsc{blue} selection and has a very similar calibrated fraction. Results are summarized in Table~\ref{tab:prior_outputs}. The tailored prior is depicted in Fig.~\ref{fig:full_result}.

\section{Properties of the photometric selections} \label{app:phot_properties}
\begin{figure}
    \centering
    \includegraphics[width=0.8\linewidth]{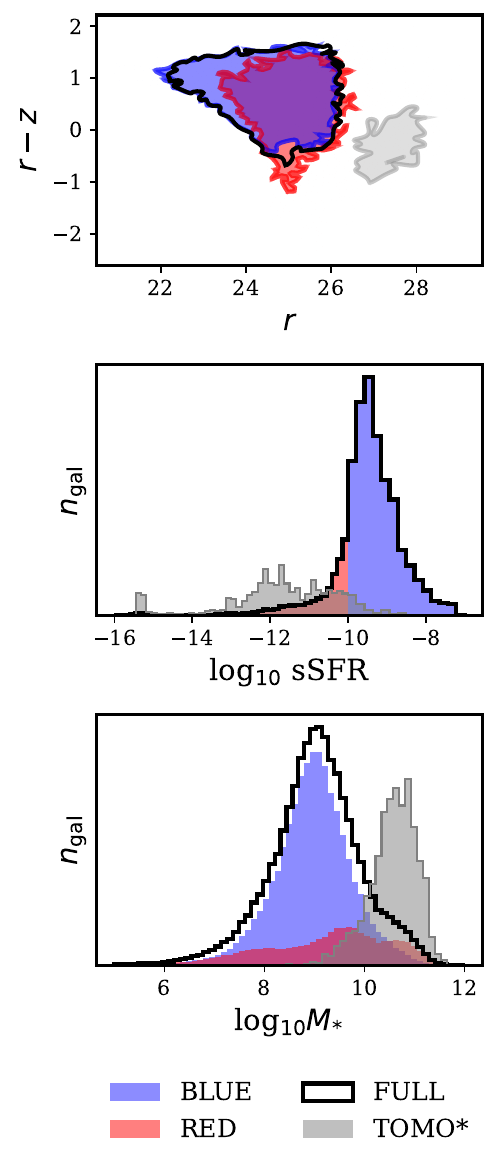}
    \caption{Distributions of properties of the photometric selections explored in the case studies of Sec.~\ref{sec:results}, including (top) 68\%-regions for DES $r-z$ color and $r$-magnitude, (center) specific star-formation rate, and (bottom) stellar mass. The bottom two panels are binned in galaxy count, with the exception of the \textsc{tomo} selection (*) which is sparser and shown with an increased relative normalization for visibility.}
    \label{fig:phot_compare}
\end{figure}

For additional information on the photometric and astrophysical properties of the samples examined in this work, we plot distributions of color and magnitude, star-formation rates, and stellar masses for the \textsc{blue}, \textsc{red}, \textsc{tomo} (gray), and \textsc{full} (black) samples in Fig.~\ref{fig:phot_compare}. It is visually apparent that the blue, star-forming sample dominates the overall population and that the \textsc{tomo} selection comprises a smaller sample of quenched, massive elliptical galaxies. 

\begin{figure}
    \centering
    \includegraphics[width=0.4\linewidth]{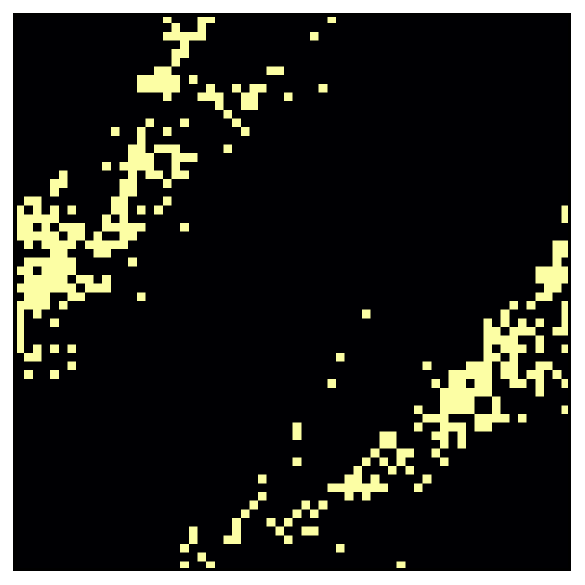}
    \caption{The deep SOM cell selection for the \textsc{tomo} sample is made up of 376 cells that prioritize low star-formation and high stellar mass. }
    \label{fig:tomomap}
\end{figure}

In Fig.~\ref{fig:tomomap}, the selected cells that constitute the \textsc{tomo} sample, detailed in Sec.~\ref{sec:tomo}, are highlighted within the analogous deep SOM (whose properties are also illustrated in Fig.~\ref{fig:allsom}).


\begin{figure}
    \centering \includegraphics[width=0.95\linewidth]{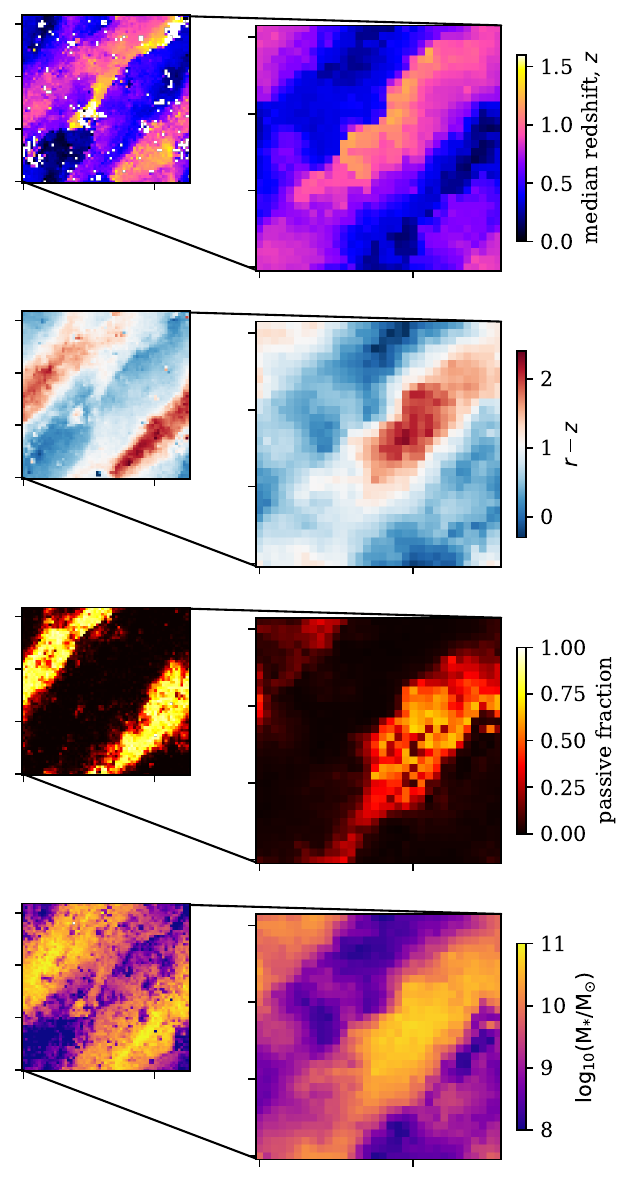}
    \caption{Self-Organizing Maps for the deep (small inset) and wide (large inset) samples of the Dark Energy Survey, and relevant astrophysical properties of each cell, from top to bottom: median redshift from spectroscopic calibrators, apparent $r-z$ color, passive fraction (meeting the \textsc{red} criteria in this work), and stellar mass.}
    \label{fig:soms_app}
\end{figure}

\section{Propagation to wide-field galaxies via \textsc{sompz}}\label{app:wide}

The Self-Organizing Map infrastructure developed and implemented in the Dark Energy Survey in \cite{buchs_2019,sanchez_2020,myles_2021,campos_2024,yin_2025}, \textsc{sompz}, is a hierarchical Bayesian inference scheme for learning the ensemble redshift distributions of a source catalog. We explore here how our tailored-alignment-prior method can leverage this structure to extend our reach to galaxy samples with poorly measured astrophysical properties. \textsc{sompz} relies on a 2-SOM approach, one that is trained on high-quality deep-field information that contains a wealth of photometric band-passes and significant overlap with spectroscopic calibration samples, and another that is trained on the limited photometry of the statistically rich shape catalog used for cosmological inference. The deep SOM contains degeneracy-breaking observational data, while the wide SOM contains the sample that we desire to calibrate. The transfer function between the two maps is learned via synthetic source injection of deep galaxies in the wide field images \citep[e.g., \textsc{balrog}:][]{balrog,y6_balrog}. 

The inference scheme for a property $a$ that is well measured in a deep field cell $c$ for a given selection of galaxies, $\hat{s}$, follows as

\begin{equation}
    p(a|b,\hat{s}) = \sum_{\hat{c}\in b}\sum_c p(a|c,\hat{s})p(c|\hat{c},\hat{s})p(\hat{c}|\hat{s}),
\end{equation}
where we define $\hat{c}$ as a cell in the wide field, and $b$ as the combination of those cells that form a given tomographic bin. Isolating each term of the inference, we see that 
\begin{itemize}
    \item $p(\hat{c}|\hat{s})$ is the overall abundance of galaxies in a given wide cell given our selection and can be written in terms of cell occupancies,
    \item $p(c|\hat{c},\hat{s})$ is our transfer function between deep and wide maps from \textsc{balrog}, and
    \item $p(a|c,\hat{s})$ is the probability distribution of $a$ from calibrators with deep field photometry -- in \textsc{sompz} this is the redshift, but for our purposes it can be alignment or the uncalibrated fraction.
\end{itemize}
There are practical adjustments that can be made to efficiently compute the above equation robustly (e.g., bin conditioning the deep-field calibrators). Example inference from the deep to wide SOM for the passive fraction of galaxies and for stellar mass can be seen in the bottom two panels of Fig.~\ref{fig:soms_app}, with the top two panels supplying the \textsc{sompz} redshift and $r-z$ color information to orient the reader. One could imagine inferring astrophysical properties for the wide SOM and using those to develop a tailored prior. However, as we care about the joint distribution of those astrophysical properties, we lose information if we perform that inference in the wide map and gain resolving power in the deep SOM.

To analogously infer an alignment prior in this system, for each deep field cell, $c$, we take the calibrators within that photometric selection and draw $p_{\mathrm{phot},c}(\mathrm{sSFR},M_*)$ to compute $f_{u,c}$. The number of calibrators that falls into each deep cell will directly impact our ability to resolve $p_{\mathrm{phot},c}(\mathrm{sSFR},M_*)$. If some cells are too sparse, they may need to be combined with their neighbors to gain statistical power. 
Rather than directly computing $p_{s,c}(A)$ from this distribution, we can simply store the overlap weights for each direct measurement, $\mathcal{O}_{i,c}$, and propagate those weights to the wide SOM as a weighted sum over all deep field cells, $c$.

\begin{equation}
    \mathcal{O}_{i,\hat{c}} = \frac{\sum_{c} w_{\hat{c}|c}\mathcal{O}_{i,c}}{\sum_{c} w_{\hat{c}|c}},
\end{equation}

where $w_{\hat{c}|c}$ is the relative contribution weight of deep cell $c$ to wide cell $\hat{c}$ learned from \textsc{balrog}.

For a given tomographic bin, $b$, we sum up the occupancy-weighted contributions of the relevant cells inside the bin, which yields for the overlap coefficients and for the uncalibrated fraction
\begin{align}
    \mathcal{O}_{i,b} &= \frac{\sum_{\hat{c}\in b} w_{\hat{c}} \mathcal{O}_{i,\hat{c}}}{\sum_{\hat{c}\in b}w_{\hat{c}}} =\frac{\sum_{\hat{c}\in b} w_{\hat{c}}\sum_{c} w_{\hat{c}|c}\mathcal{O}_{i,c}}{\sum_{\hat{c}\in b}w_{\hat{c}}\sum_{c} w_{\hat{c}|c}}, \ \ \text{and}\\
    f_{u,b} &= \frac{\sum_{\hat{c}\in b} w_{\hat{c}} {f}_{u,\hat{c}}}{\sum_{\hat{c}\in b}w_{\hat{c}}} =\frac{\sum_{\hat{c}\in b} w_{\hat{c}}\sum_{c} w_{\hat{c}|c}{f}_{u,c}}{\sum_{\hat{c}\in b}w_{\hat{c}}\sum_{c} w_{\hat{c}|c}},
\end{align}
where $w_{\hat{c}}$ is the abundance weight of wide cell $\hat{c}$. Using these inferred ingredients for each tomographic bin selection, we can then propagate to our final results for using Eqns.~5-10 in Sec.~\ref{sec:methods}. If statistics allow, additional selections on the galaxies used to inform $\mathcal{O}_{i,c},\ f_{u,c}$ can be imposed to limit the measurement within the deep field to galaxies that would be selected in the wide tomographic bin, thus bin-conditioning the calibrators for the tailored prior as well as the photometric redshift inference.

As this method and \textsc{sompz} rely on deep calibrating information often in the same fields, characterization of systematic uncertainty by perturbing deep field calibrators \citep[e.g.,][]{yin_2025} could similarly be extended to the tailored-IA method inputs to simultaneously constrain both and produce correlated redshift-alignment priors. This is left to future work.


\bsp	
\label{lastpage}
\end{document}